# A New Decomposition Strategy for Analyzing Large-Scale Systems Such as Power Systems

Minquan Chen, Deqiang Gan, *Senior Member*, *IEEE*

***Abstract*—In this work, it is demonstrated that the usual power system dynamic model exhibits a feedforward-feedback control structure. The distinct properties of the feedforward and feedback subsystems are identified and studied using respective nonlinear system tools. The stability of the closed-loop system is investigated using a small gain argument from input-to-state stability theory. Test results are provided to further complement the theoretical findings. The introduced methodology also shed light on the dynamics study on other interconnected nonlinear systems.**

***Index Terms* — feedforward-feedback, decomposition-aggregation, Kuramoto coupling oscillator model, monotone system, input to state stability, power system.**

## I. Introduction

THE subject of power system dynamics has a very long and rich history with extensive literature [1], [2]. As a power system model is typically high-dimensional and nonlinear, dynamics analysis of power systems presents itself as a major outstanding challenge to power community.

Many methods have been proposed in the past decades to cope with the above notably complex engineering problem. While numerical integration continues to be widely accepted by the industry [3], energy function methods [4] - [6] and functional approximation [7] - [10] have also received much attention. In recent years, asymptotic expansion methods are also gaining popularity and advancing rather rapidly [11] - [13].

Most of the results mentioned above are intended to be computational rather than analytical. Otherwise, the present work concerns the structural properties of power system dynamics. It is shown that the subsystems of a power system model exhibit distinct characteristics, which indicates that a decomposition-aggregation approach is naturally suitable for attacking a theoretical challenge like power system dynamics.

The notion of decomposition-aggregation is not new. It had been proposed in [14] - [17] to reduce computational burden. This decomposition and aggregation strategy proceeds by dividing a large-scale system into smaller *control areas* and iteratively executing subsequent simulations. In the present work, we revisit this divide-and-conquer paradigm [18], presenting a fundamentally different implementation strategy: decomposing the original system model based on the *natures* of the subsystems. This new strategy allows one to gain much insight into the structural properties of a power system model.

This work was supported by Natural Science Foundation of China (Transient stability analysis and control of AC/DC power grids with high penetration of renewable power generation, U2166601). *(Corresponding author: D. Gan).*

Minquan Chen and Deqiang Gan are with the School of Electrical Engineering, Zhejiang University, Hangzhou, China. (e-mail: ee_cmq@zju.edu.cn; deqiang.gan@ieee.org).

The rest of this paper is organized as follows: Section II proposes a closed-loop control structure for a power system dynamic model. Section III focuses on the feedforward subsystem's stability characteristics, which plays an important role in rotor angle stability. In Section IV, we study in detail the properties of the feedback subsystem which has a dominant impact on voltage behavior. Having studied the properties of feedforward and feedback subsystems, the stability of the closed-loop system is investigated in Section V. Section VI reports upon several case studies and Section VII provides a summary and a discussion of future research. Finally, the Appendix section describes how to extend the introduced results to tackle more complex dynamic models.

## II. Feedforward-Feedback Control Structure of Power System Models

Consider a power system with $n$ generating units coupled via a connected transmission network. Its dynamics can be described by the following differential-algebraic equations (DAE) model:

$$\begin{cases} \dot{\boldsymbol{x}} = \boldsymbol{f}(\boldsymbol{x}, \boldsymbol{y}) \\ \boldsymbol{0} = \boldsymbol{g}(\boldsymbol{x}, \boldsymbol{y}) \end{cases},$$

where $\boldsymbol{x}$, $\boldsymbol{y}$ are the vectors of state variables and algebraic variables, respectively.

If synchronous machines rotors adopt a one-axis model and exciters adopt a first-order model, the system dynamics are expressed as:

$$\begin{cases} \dot{\boldsymbol{\delta}} = \omega_0(\boldsymbol{\omega} - \boldsymbol{\omega}_{\text{ref}}) \\ 2\boldsymbol{H}\dot{\boldsymbol{\omega}} = \boldsymbol{P}_{\text{m}} - \boldsymbol{I}_q \circ \boldsymbol{E}'_q - (\boldsymbol{X}'_q - \boldsymbol{X}'_d)(\boldsymbol{I}_d \circ \boldsymbol{I}_q) - \boldsymbol{D}\boldsymbol{\omega} \\ \boldsymbol{T}'_{d0}\dot{\boldsymbol{E}}'_q = \boldsymbol{E}_{\text{fd}} - \boldsymbol{E}'_q - (\boldsymbol{X}_d - \boldsymbol{X}'_d)\boldsymbol{I}_d \\ \boldsymbol{T}_{\text{A}}\dot{\boldsymbol{E}}_{\text{fd}} = \boldsymbol{K}_{\text{A}}(\boldsymbol{V}_{\text{ref}} - |\boldsymbol{V}_{\text{G}}|) - \boldsymbol{E}_{\text{fd}} \end{cases}, \quad (1)$$

where operator $\circ$ indicates Hadamard product.

The state variables of each generator consist of rotor angle $\delta$, rotor speed $\omega$, $q$-axis transient voltage $E'_q$ and excitation voltage $E_{\text{fd}}$. For the definitions of the other variables and parameters, the readers are referred to [2].

Define $\boldsymbol{Y}_{\text{GG}}$, $\boldsymbol{Y}_{\text{GL}}$, $\boldsymbol{Y}_{\text{LG}}$, $\boldsymbol{Y}_{\text{LL}}$ as the submatrices of the network admittance matrix. Further, by summarizing all loads in a diagonal admittance matrix $\boldsymbol{Y}_{\text{L}}$, the following network equation can be obtained:

$$\begin{cases} \boldsymbol{Y}_{\text{GG}}\boldsymbol{V}_{\text{G}} + \boldsymbol{Y}_{\text{GL}}\boldsymbol{V}_{\text{L}} = -\text{j}\boldsymbol{X}'^{-1}_d e^{\text{j}\boldsymbol{\delta}}\boldsymbol{E}'_q - \text{j}0.5(\boldsymbol{X}'^{-1}_q - \boldsymbol{X}'^{-1}_d)e^{\text{j}2\boldsymbol{\delta}}\boldsymbol{V}_{\text{G}} \\ \boldsymbol{Y}_{\text{LG}}\boldsymbol{V}_{\text{G}} + \boldsymbol{Y}_{\text{LL}}\boldsymbol{V}_{\text{L}} = \boldsymbol{Y}_{\text{L}}\boldsymbol{V}_{\text{L}} \end{cases}.$$

For simplicity and without loss of representativeness, let $X'_d = X'_q$. After eliminating load voltage $V_L$ using Kron reduction, one obtains the expressions of the injection currents of synchronous machines as follows [2]:

$$\begin{cases} \boldsymbol{I}_d = \underbrace{(\boldsymbol{G}\circ\sin\boldsymbol{\delta}_\Delta - \boldsymbol{B}\circ\cos\boldsymbol{\delta}_\Delta)}_{\boldsymbol{K}_d}\boldsymbol{E}'_q = \boldsymbol{K}_d\boldsymbol{E}'_q \\ \boldsymbol{I}_q = \underbrace{(\boldsymbol{G}\circ\cos\boldsymbol{\delta}_\Delta + \boldsymbol{B}\circ\sin\boldsymbol{\delta}_\Delta)}_{\boldsymbol{K}_q}\boldsymbol{E}'_q = \boldsymbol{K}_q\boldsymbol{E}'_q \end{cases}, \quad (2)$$

$$\boldsymbol{\delta}_\Delta = \begin{bmatrix} \delta_1-\delta_1 & \cdots & \delta_1-\delta_n \\ \vdots & \ddots & \vdots \\ \delta_n-\delta_1 & \vdots & \delta_n-\delta_n \end{bmatrix}, \quad (3)$$

where $\boldsymbol{G}$, $\boldsymbol{B}$ denote the real and imaginary parts of the reduced-order network admittance matrix, respectively.

As a result, the DAE model is simplified into a set of pure ordinary differential equations as follows:

$$\begin{cases} \dot{\boldsymbol{\delta}} = \omega_s(\boldsymbol{\omega}-\boldsymbol{\omega}_{\text{ref}}) \\ 2\boldsymbol{H}\dot{\boldsymbol{\omega}} = \boldsymbol{P}_{\text{m}} - \boldsymbol{K}_q\boldsymbol{E}'_q\circ\boldsymbol{E}'_q - \boldsymbol{D}\boldsymbol{\omega} \\ \boldsymbol{T}'_{d0}\dot{\boldsymbol{E}}'_q = \boldsymbol{E}_{\text{fd}} - \boldsymbol{E}'_q - (\boldsymbol{X}_d-\boldsymbol{X}'_d)\boldsymbol{K}_d\boldsymbol{E}'_q \\ \boldsymbol{T}_{\text{A}}\dot{\boldsymbol{E}}_{\text{fd}} = \boldsymbol{K}_{\text{A}}\boldsymbol{V}_{\text{ref}} - \boldsymbol{E}_{\text{fd}} - \boldsymbol{K}_{\text{A}}|\boldsymbol{V}_{\text{G}}| \end{cases}. \quad (4)$$

It should be noted that the results to be presented can also be generalized to deal with DAE models. The appendix provides some preliminary results in this direction.

Despite its popularity, dynamic model (2) - (4) is so complicated that it is typically studied using a numerical simulation approach. To gain insights into power system dynamics, a new decomposition-aggregation strategy is suggested below. Consider the rotor dynamics subsystem:

$$\begin{cases} \dot{\boldsymbol{\delta}} = \omega_s(\boldsymbol{\omega}-\boldsymbol{\omega}_{\text{ref}}) \\ 2\boldsymbol{H}\dot{\boldsymbol{\omega}} = \boldsymbol{P}_{\text{m}} - \boldsymbol{K}_q\boldsymbol{E}'_q\circ\boldsymbol{E}'_q - \boldsymbol{D}\boldsymbol{\omega} \end{cases}. \quad (5)$$

Let us temporarily fix $q$-axis transient voltage $\boldsymbol{E}'_q$, then it will be shortly shown that the above rotor subsystem can be viewed as a second-order Kuramoto coupling oscillator model (Kuramoto model for short) [19], [20]. Alternatively, the rotor subsystem can be viewed as a Kuramoto model with input $\boldsymbol{E}'_q$.

By the same token, rotor angle $\boldsymbol{\delta}$ can be viewed as the input to the following voltage dynamics subsystem:

$$\begin{cases} \boldsymbol{T}'_{d0}\dot{\boldsymbol{E}}'_q = \boldsymbol{E}_{\text{fd}} - \boldsymbol{E}'_q - (\boldsymbol{X}_d-\boldsymbol{X}'_d)\boldsymbol{K}_d\boldsymbol{E}'_q \\ \boldsymbol{T}_{\text{A}}\dot{\boldsymbol{E}}_{\text{fd}} = \boldsymbol{K}_{\text{A}}\boldsymbol{V}_{\text{ref}} - \boldsymbol{E}_{\text{fd}} - \boldsymbol{K}_{\text{A}}|\boldsymbol{V}_{\text{G}}| \end{cases}. \quad (6)$$

One can go one step further, by decomposing the above voltage subsystem into $\boldsymbol{E}'_q$ subsystem and $\boldsymbol{E}_{fd}$ subsystem. The significance of this last decomposition will become clear in Section IV, where it will be shown that both $\boldsymbol{E}'_q$ and $\boldsymbol{E}_{fd}$ subsystems are the so-called monotone systems.

The above arguments indicate immediately that the dynamic model (4) demonstrates a feedforward-feedback control structure, as Fig. 1 depicts. The natures of the feedforward and feedback subsystems are rather different. It is not difficult to understand that, this decomposition-aggregation strategy permits us to gain insights into the ever-increasing complex dynamics of power systems.

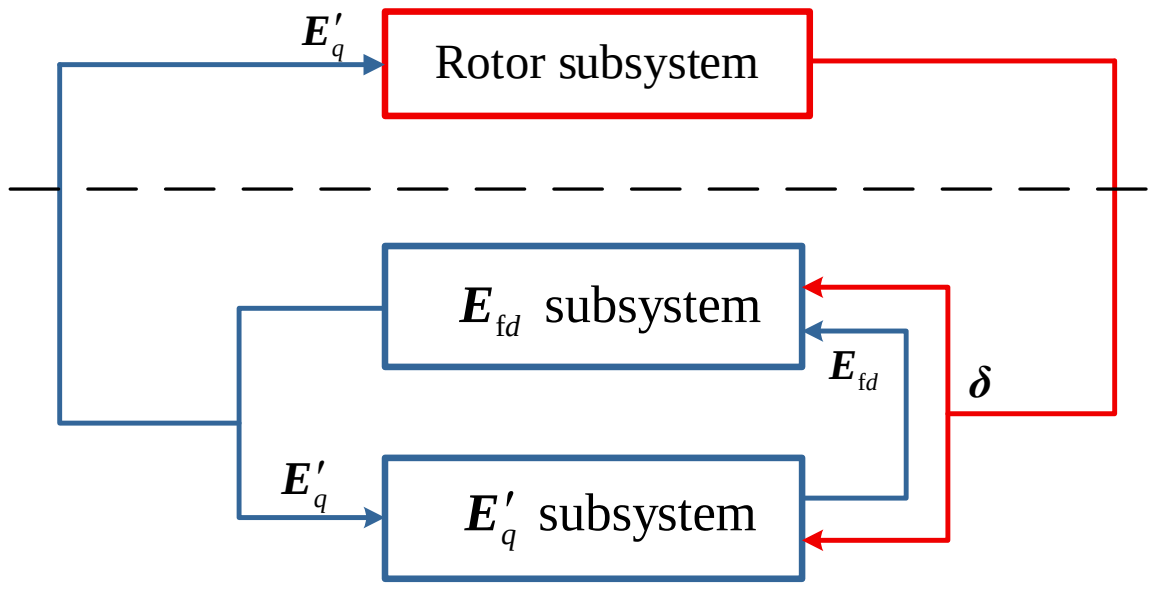

Fig. 1. Feedforward-feedback structure of power system model.

In the subsequent text, the unique properties of the feedforward and feedback subsystems will be discussed in Section III and Section IV, respectively. The stability of the closed-loop system will be addressed in Section V.

## III. Properties of Feedforward Subsystem

For completeness, in this section we briefly review the relevant properties of the feedforward subsystem.

Mark $\boldsymbol{Y} = \boldsymbol{G} + \mathrm{j}\boldsymbol{B}$ and $\alpha_{ij} = \angle Y_{ij} - \pi/2$, thus introducing $K_{ij} = |E'_{qi}E'_{qj}Y_{ij}|$ and $\Omega_{ij} = P_{\text{m}i} - D_i\omega_{\text{ref}} - E'_{qi}E'_{qj}G_{ij}$. Then an alternative formulation of the feedforward subsystem can be obtained:

$$2H_i\ddot{\delta}_i + D_i\dot{\delta}_i/\omega_{\text{ref}} = \Omega_i + \sum\nolimits_{j=1,j\neq i}^{n} K_{ij}\sin(\delta_j-\delta_i+\alpha_{ij}). \quad (7)$$

There exists extensive literature on the stability margin estimation of (5) or (7), notably [4] – [6] among others. The analytical expression of stability region boundaries of the above system is addressed in [21]. Recently, it was found that there is striking similarity between the above model and the much-studied Kuramoto model. A recent result, central in modern Kuramoto regime, is the following explicit formula for the estimation of stability region [22], [23]:

$$L\sin\Theta_{\text{c}} > \Theta(\boldsymbol{\Omega}),$$

where $L$ and $\Theta(\boldsymbol{\Omega})$ are explicitly defined in terms of $K_{ij}, H_i$ etc.; $\Theta_{\text{c}}$ is an implicit function of $\boldsymbol{\Omega}$, inertia $\boldsymbol{H}$ and damping constant $\boldsymbol{D}$. For results of similar flavor, the readers are referred to [24] for details.

It is well-understood that the feedforward system (7) is linearly stable as long as the damping coefficients are positive (which is typically the case). The linear stability is also related to the algebraic connectivity of the Laplacian graph defined by coupling coefficients $K_{ij}$. There is also explicit relationship between algebraic connectivity and the existence of equilibrium point. We will not pursue this issue further, and reference [25] presents the latest development in this line of research.

## IV. Properties of Feedback Subsystem

The properties of the feedback subsystem, which apparently play a dominant role in power system voltage dynamics, are examined in detail in this section. While the rationale a modern power system regulates voltage is well-understood, the theoretical foundation of this wisdom is in fact not known. The results of this section provide partial answers to this important question. The readers are referred to [26] for some early attempts in this direction.

### A. Sign pattern of Jacobian matrix

The key in understanding the feedback subsystem is to recognize the inherent sign pattern of the subsystem Jacobian matrix. Let $\mathbf{I}_n$ be an $n\times n$ unit matrix, it is not difficult to see that:

$$\boldsymbol{T}'_{d0}\dot{\boldsymbol{E}}'_q = \boldsymbol{E}_{\mathrm{fd}} - \underbrace{[\mathbf{I}_n + (\boldsymbol{X}_d - \boldsymbol{X}'_d)\boldsymbol{K}_d]}_{\boldsymbol{K}_{\mathrm{V}}}\boldsymbol{E}'_q = \boldsymbol{E}_{\mathrm{fd}} + \boldsymbol{K}_{\mathrm{V}}\boldsymbol{E}'_q .$$

For voltage dynamics analysis, the following assumption is often used, which holds true when rotor angles do not swing against each other radically:

$$\left|\boldsymbol{G}\circ \sin\boldsymbol{\delta}_\Delta\right| << \left|\boldsymbol{B}\circ \cos\boldsymbol{\delta}_\Delta\right| . \tag{8}$$

Since matrix $\boldsymbol{B}$ is approximately a loopy Laplacian matrix [27], the above inequality has very important consequences, and it will be summarized later in Assertion 1. Recognizing (8), $\boldsymbol{K}_d \approx -\boldsymbol{B}\circ\cos\delta_\Delta$ can be obtained, which is a structured matrix. For instance, the numerical results of $\boldsymbol{K}_d$ and $\boldsymbol{K}_{\mathrm{V}}$ of a 3-machine test system (see [28] for data) are shown below:

$$\boldsymbol{K}_d = \begin{bmatrix} 1.735 & -0.55 & -0.42 \\ -0.07 & 0.861 & -0.10 \\ -0.11 & -0.11 & 0.644 \end{bmatrix},$$

$$\boldsymbol{K}_{\mathrm{V}} = \begin{bmatrix} -1.09 & 0.027 & 0.021 \\ 0.002 & -1.03 & 0.003 \\ 0.006 & 0.006 & -1.04 \end{bmatrix}.$$

From the complex form expression of terminal voltage:

$$\begin{aligned}\bar{\boldsymbol{V}}_{\mathrm{G}} &= [\mathrm{j}\mathbf{I}_n - \mathrm{j}\boldsymbol{X}'_d(\boldsymbol{K}_d + \mathrm{j}\boldsymbol{K}_q)]\boldsymbol{E}'_q = \\ &[\underbrace{\boldsymbol{X}'_d\boldsymbol{K}_q}_{\boldsymbol{K}_{\mathrm{R}}} + \mathrm{j}\underbrace{(\mathbf{I}_n - \boldsymbol{X}'\boldsymbol{K}_d)}_{\boldsymbol{K}_{\mathrm{I}}}]\boldsymbol{E}'_q = (\boldsymbol{K}_{\mathrm{R}} + \mathrm{j}\boldsymbol{K}_{\mathrm{I}})\boldsymbol{E}'_q,\end{aligned}$$

it follows that the terminal voltage of the *i-th* generator ($|V_{\mathrm{G}i}|$) can be expressed as

$$|V_{\mathrm{G}i}| = h_i(\boldsymbol{E}'_q) = \sqrt{\boldsymbol{E}'^{\mathrm{T}}_q \boldsymbol{C}_i \boldsymbol{E}'_q} = \sqrt{\sum_{j=1}^{n}\sum_{k=1}^{n} C_{i,jk} E'_{q,j} E'_{q,k}} ,$$

where $C_{i,jk} = K_{\mathrm{R}ij}K_{\mathrm{R}ik} + K_{\mathrm{I}ij}K_{\mathrm{I}ik}$ are coefficient matrices determined by $\boldsymbol{K}_{\mathrm{R}}$ and $\boldsymbol{K}_{\mathrm{I}}$; $h_i(\cdot)$ denotes a nonlinear function. Matrices $\boldsymbol{K}_{\mathrm{R}}$ and $\boldsymbol{K}_{\mathrm{I}}$ are also highly structured. Here is an example obtained from the 3-machine system:

$$\begin{cases} \boldsymbol{K}_{\mathrm{R}} = \begin{bmatrix} 0.140 & -0.03 & -0.02 \\ 0.557 & 0.075 & 0.059 \\ 0.572 & 0.052 & 0.056 \end{bmatrix} \\ \boldsymbol{K}_{\mathrm{I}} = \begin{bmatrix} 0.832 & 0.05 & 0.041 \\ 0.063 & 0.256 & 0.087 \\ 0.145 & 0.143 & 0.190 \end{bmatrix} \end{cases}.$$

As we can see from the example that $\boldsymbol{K}_{\mathrm{I}}$ is a positive matrix while $\boldsymbol{K}_{\mathrm{R}}$ has only several small negative entries. Afterward, the coefficient matrices $\boldsymbol{C}_i$ can be determined as follows:

$$\boldsymbol{C}_1 = \begin{bmatrix} 0.712 & 0.040 & 0.031 \\ 0.040 & 0.004 & 0.003 \\ 0.031 & 0.003 & 0.002 \end{bmatrix}.$$

From the numerical examples of $\boldsymbol{K}_{\mathrm{R}}$ and $\boldsymbol{K}_{\mathrm{I}}$, it is not difficult to determine that each entry of $\boldsymbol{C}_i$ is non-negative.

**Assertion 1** Matrices $\boldsymbol{K}_d$ and $-\boldsymbol{K}_{\mathrm{V}}$ are diagonally dominant with positive diagonal entries and non-positive off-diagonal entries; Matrices $\boldsymbol{C}_i$ ($i$= 1, …, $n$) are non-negative; the Jacobian matrix $\partial\boldsymbol{h}/\partial\boldsymbol{E}'_q \geq \boldsymbol{0}$ .

Let us now re-write the feedback subsystem as follows:

$$\begin{bmatrix} \boldsymbol{T}'_{d0}\dot{\boldsymbol{E}}'_q \\ \boldsymbol{T}_{\mathrm{A}}\dot{\boldsymbol{E}}_{\mathrm{fd}} \end{bmatrix} = \begin{bmatrix} \boldsymbol{K}_{\mathrm{V}} & \mathbf{I}_n \\ \mathbf{0} & -\mathbf{I}_n \end{bmatrix}\begin{bmatrix} \boldsymbol{E}'_q \\ \boldsymbol{E}_{\mathrm{fd}} \end{bmatrix} + \begin{bmatrix} \mathbf{0} \\ \boldsymbol{K}_{\mathrm{A}}(\boldsymbol{V}_{\mathrm{ref}} - \boldsymbol{h}(\boldsymbol{E}'_q)) \end{bmatrix}. \tag{9}$$

One of the most pleasant findings in this work is that the Jacobian matrix $\boldsymbol{J}$ of the above subsystem exhibits a stable sign pattern. Take the Jacobian matrix of the voltage subsystem of a 3-machine test system as an example, as shown below:

$$\mathrm{sgn}(\boldsymbol{J}) = \mathrm{sgn}\begin{pmatrix} \dfrac{\partial\dot{\boldsymbol{E}}'_q}{\partial\boldsymbol{E}'_q} & \dfrac{\partial\dot{\boldsymbol{E}}'_q}{\partial\boldsymbol{E}_{\mathrm{fd}}} \\ \dfrac{\partial\dot{\boldsymbol{E}}_{\mathrm{fd}}}{\partial\boldsymbol{E}'_q} & \dfrac{\partial\dot{\boldsymbol{E}}_{\mathrm{fd}}}{\partial\boldsymbol{E}_{\mathrm{fd}}} \end{pmatrix} = \begin{bmatrix} - & + & + & + & 0 & 0 \\ + & - & + & 0 & + & 0 \\ + & + & - & 0 & 0 & + \\ - & - & - & - & 0 & 0 \\ - & - & - & 0 & - & 0 \\ - & - & - & 0 & 0 & - \end{bmatrix}, \tag{10}$$

where sgn(·) denotes the sign result; the symbol '+' ('−', '0') indicates that the entry is non-negative (non-positive, zero).

The sign pattern reported above shows that each entry of the Jacobian matrix is sign stable, indicating that the interactions between state variables are robust and well-behaved. For instance, a rise in one transient voltage will increase all the other transient voltages but decrease all excitation voltages. Besides, all negative diagonal entries are conducive to the stability of the voltage dynamics. This explains how multiple synchronous machines in a modern power system maintain voltage profiles in a cooperative yet decentralized way.

The sign pattern of Jacobian matrix $\boldsymbol{J}$ can be further probed, the results are presented in Section IV-B and IV-C subsequently.

### B. Positively Invariant Set of the Subsystem

Consider an autonomous system governed by the following differential equations:

$$\dot{\boldsymbol{x}} = \boldsymbol{f}(\boldsymbol{x}), \boldsymbol{x} \in \mathcal{X}$$

where state space $\mathcal{X} \subseteq \mathbb{R}^n$ and $\boldsymbol{f}(\boldsymbol{x})$: $\mathcal{X} \to \mathbb{R}^n$ is differentiable.

If its Jacobian matrix $\boldsymbol{J}(\boldsymbol{x})$ is a Metzler matrix (every non-diagonal element is non-negative) for all $\boldsymbol{x} \in \mathcal{X}$, then the system is *monotone*. A pleasant property of a monotone system is that all state variables reinforce each other, so the system effectively behaves like a scalar system. Besides, monotone systems are globally stable, almost always.

As we showed in Section IV.A, under mild conditions, the Jacobian matrix of the feedback system is sign stable. Such systems are called *mixed monotone*. The sign stability of $\boldsymbol{J}(\boldsymbol{x})$ is equivalent to the existence of a matrix $\boldsymbol{P} \subseteq \mathbb{R}^{n\times n}$ and a function $\boldsymbol{F}$: $\mathcal{X}\times\mathcal{X} \to \mathbb{R}^n$ satisfying the following conditions:

1) $\forall \boldsymbol{x} \in \mathcal{X} \Rightarrow \boldsymbol{F}(\boldsymbol{x},\boldsymbol{x}) = \boldsymbol{f}(\boldsymbol{x})$,
2) $\forall \boldsymbol{x}^-, \boldsymbol{x}^+, \boldsymbol{y} \in \mathcal{X}, \boldsymbol{Px}^- \leq \boldsymbol{Px}^+ \Rightarrow \boldsymbol{PF}(\boldsymbol{x}^-,\boldsymbol{y}) \leq \boldsymbol{PF}(\boldsymbol{x}^+,\boldsymbol{y})$,
3) $\forall \boldsymbol{x}, \boldsymbol{y}^-, \boldsymbol{y}^+ \in \mathcal{X}, \boldsymbol{Py}^- \leq \boldsymbol{Py}^+ \Rightarrow \boldsymbol{PF}(\boldsymbol{x},\boldsymbol{y}^-) \geq \boldsymbol{PF}(\boldsymbol{x},\boldsymbol{y}^+)$.

The function $\boldsymbol{F}$ is isotonic in terms of the first argument and antitonic in terms of the second argument. As shown in [29], the trajectories of the following augmented system form (see Fig. 2) envelope of the trajectories of the original system:

$$\begin{cases} \dot{\boldsymbol{x}}^+ = \boldsymbol{F}(\boldsymbol{x}^+, \boldsymbol{x}^-) \\ \dot{\boldsymbol{x}}^- = \boldsymbol{F}(\boldsymbol{x}^-, \boldsymbol{x}^+) \end{cases},$$

where $\boldsymbol{x}^+, \boldsymbol{x}^- \in \mathcal{X}$.

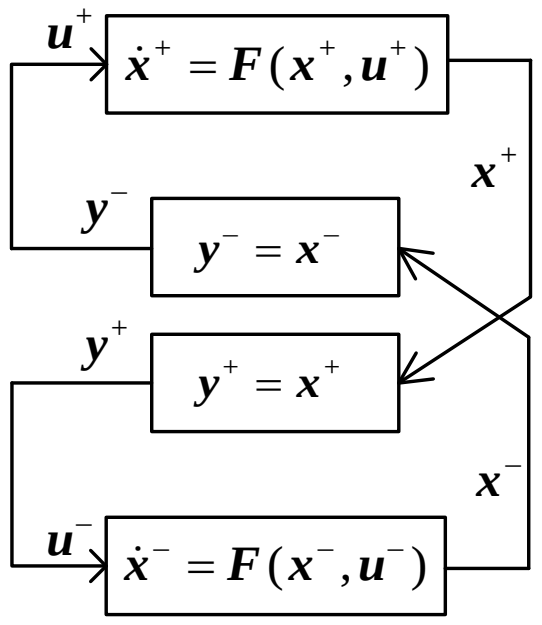

Fig. 2. Structure of augmented system.

The above augmented system should permit us to infer the quantitative result of the original system, as the following result confirms.

**Theorem 1**[30]. Suppose there exists states $\boldsymbol{x}^- < \boldsymbol{x}^+$ such that $\boldsymbol{F}(\boldsymbol{x}^+, \boldsymbol{x}^-) \leq \boldsymbol{0} \leq \boldsymbol{F}(\boldsymbol{x}^-, \boldsymbol{x}^+)$, then $[\boldsymbol{x}^-, \boldsymbol{x}^+]$ is positively invariant for the system $\dot{\boldsymbol{x}} = \boldsymbol{f}(\boldsymbol{x}), \boldsymbol{x} \in \mathcal{X}$.

For instance, consider the following dynamic system:

$$\begin{cases} \dot{x}_1 = -2x_1 + x_2 \\ \dot{x}_2 = 10 - 10x_1 - 10x_2 \end{cases}.$$

Obviously, it is mixed monotone so that its augmented system can be constructed as:

$$\begin{cases} \dot{x}_1^+ = -2x_1^+ + x_2^+ \\ \dot{x}_2^+ = 10 - 10x_1^- - 10x_2^+ \\ \dot{x}_1^- = -2x_1^- + x_2^- \\ \dot{x}_2^- = 10 - 10x_1^+ - 10x_2^- \end{cases}.$$

We find that $[x_1^+, x_2^+] = [0.8, 1.1]$ and $[x_1^-, x_2^-] = [0.0, 0.1]$ satisfy the conditions listed in Theorem 1. As shown in Fig. 3, none of the trajectories (solid lines) within this set drift out.

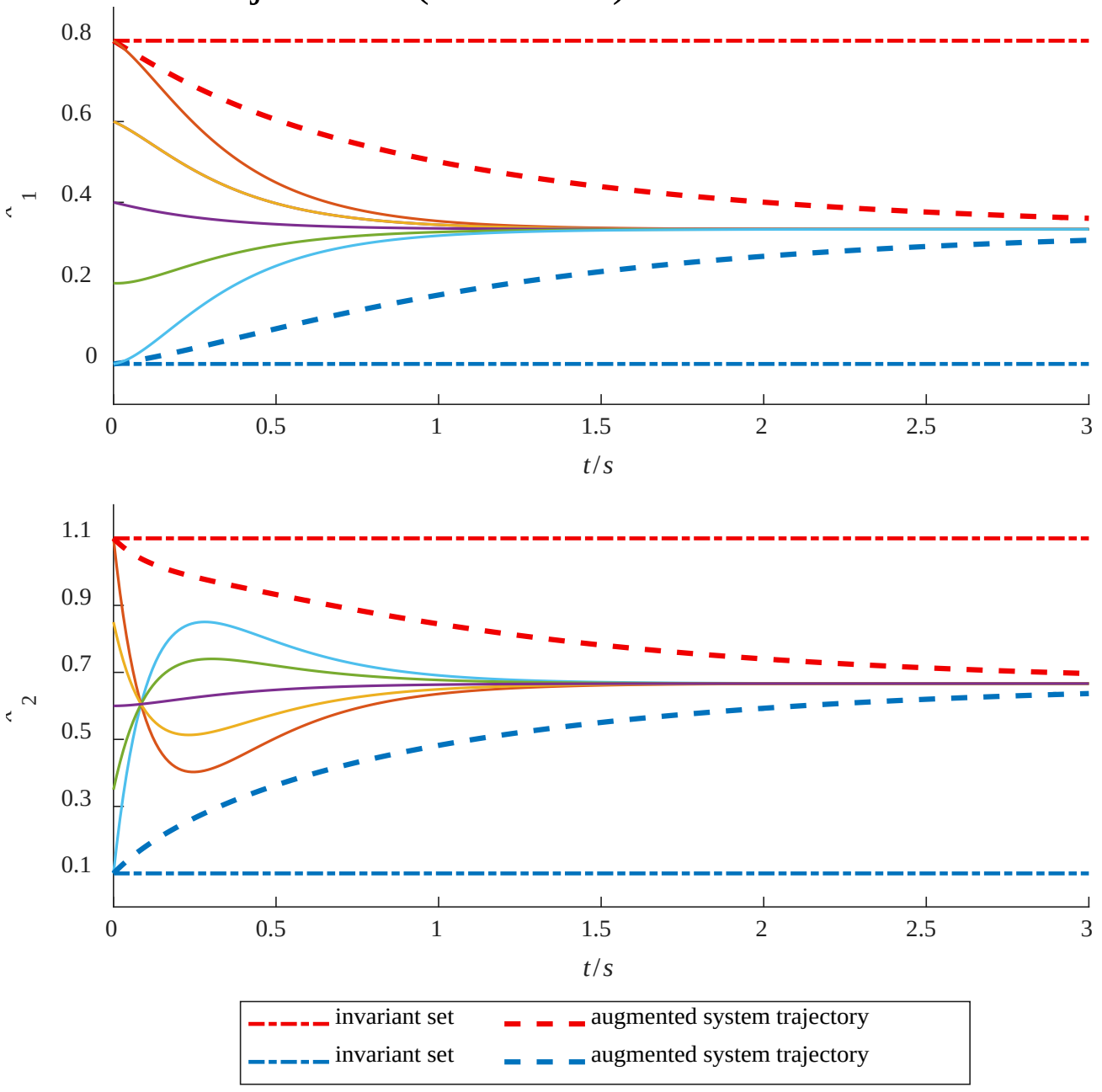

Fig. 3. Time-domain trajectories and invariant set.

Based on the sign pattern reported in (10), one constructs easily the following augmented system to analyze voltage stability. For example, if we add a 100MVar reactive assumption at Bus-5 in the 3-machine test system, the nodal voltages will fluctuate. Using Theorem 1, an invariant set (see Fig. 4) can be obtained from the following augmented system:

$$\begin{bmatrix} \boldsymbol{T}'_{d0}\dot{\boldsymbol{E}}'^{+}_{q} \\ \boldsymbol{T}_{\mathrm{A}}\dot{\boldsymbol{E}}^{+}_{\mathrm{fd}} \\ \boldsymbol{T}'_{d0}\dot{\boldsymbol{E}}'^{-}_{q} \\ \boldsymbol{T}_{\mathrm{A}}\dot{\boldsymbol{E}}^{-}_{\mathrm{fd}} \end{bmatrix} = \begin{bmatrix} \boldsymbol{K}_{\mathrm{V}} & \mathbf{I}_n & \mathbf{0} & \mathbf{0} \\ \mathbf{0} & -\mathbf{I}_n & \mathbf{0} & \mathbf{0} \\ \mathbf{0} & \mathbf{0} & \boldsymbol{K}_{\mathrm{V}} & \mathbf{I}_n \\ \mathbf{0} & \mathbf{0} & \mathbf{0} & -\mathbf{I}_n \end{bmatrix} \begin{bmatrix} \boldsymbol{E}'^{+}_{q} \\ \boldsymbol{E}^{+}_{\mathrm{fd}} \\ \boldsymbol{E}'^{-}_{q} \\ \boldsymbol{E}^{-}_{\mathrm{fd}} \end{bmatrix} + \begin{bmatrix} \mathbf{0} \\ \boldsymbol{K}_{\mathrm{A}}(\boldsymbol{V}_{\mathrm{ref}} - \boldsymbol{h}(\boldsymbol{E}'^{-}_{q})) \\ \mathbf{0} \\ \boldsymbol{K}_{\mathrm{A}}(\boldsymbol{V}_{\mathrm{ref}} - \boldsymbol{h}(\boldsymbol{E}'^{+}_{q})) \end{bmatrix}$$

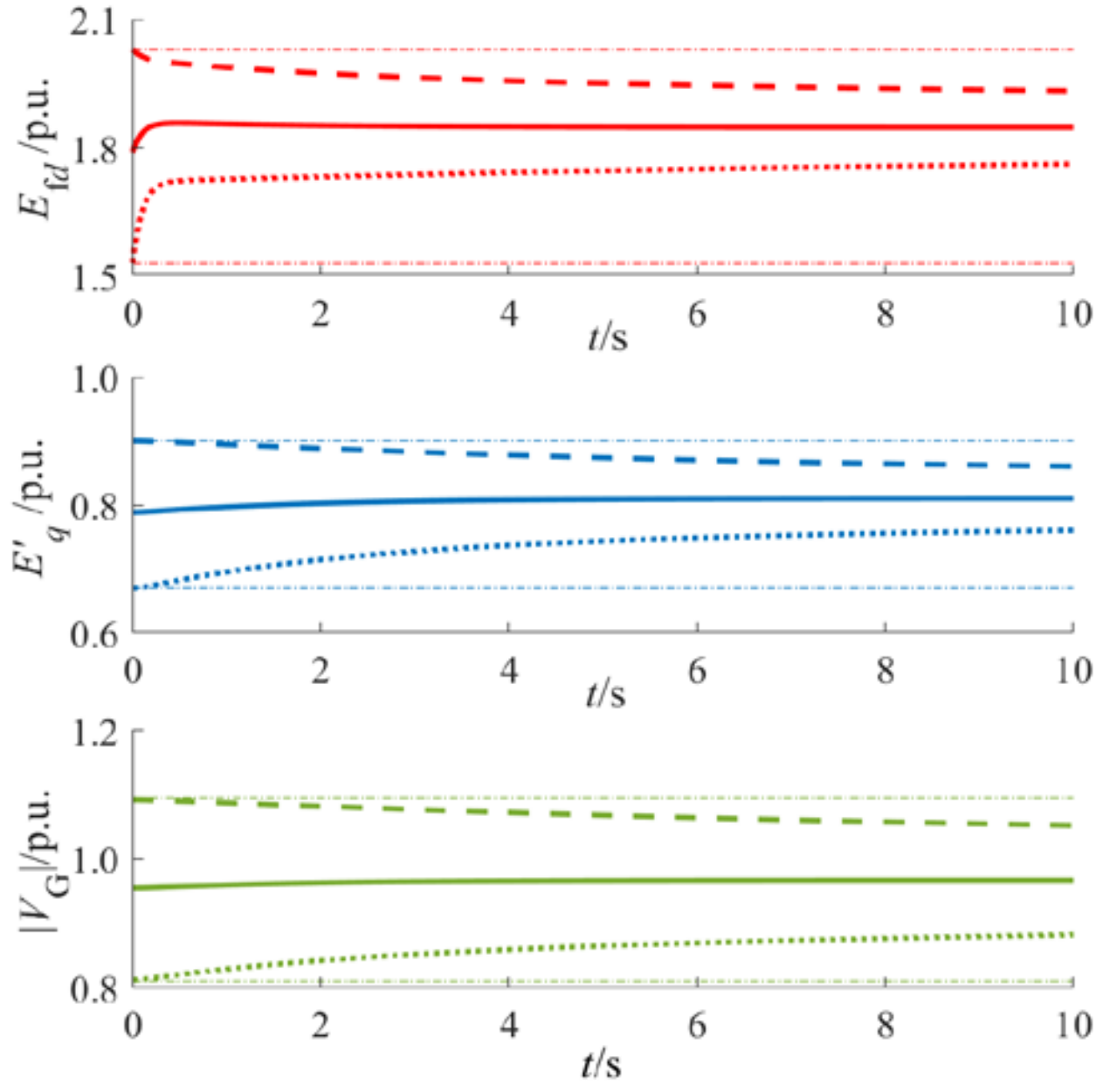

Fig. 4. Invariant set of $G_2$ in 3-machine test system.

Consider now the effect of the excitation saturation functions: $\boldsymbol{E}_{\mathrm{fd,min}} \leq \mathrm{Sat}(\boldsymbol{E}_{\mathrm{fd}}) \leq \boldsymbol{E}_{\mathrm{fd,min}}$. Since Sat($\boldsymbol{E}_{\mathrm{fd}}$) is also monotone, the monotonicity of the augmented system will not be broken. Fig. 5 compares the difference in voltage response before and after adding the saturation function Sat($\boldsymbol{E}_{\mathrm{fd}}$). Such a control design is conducive to reducing the fluctuation of the voltage responses.

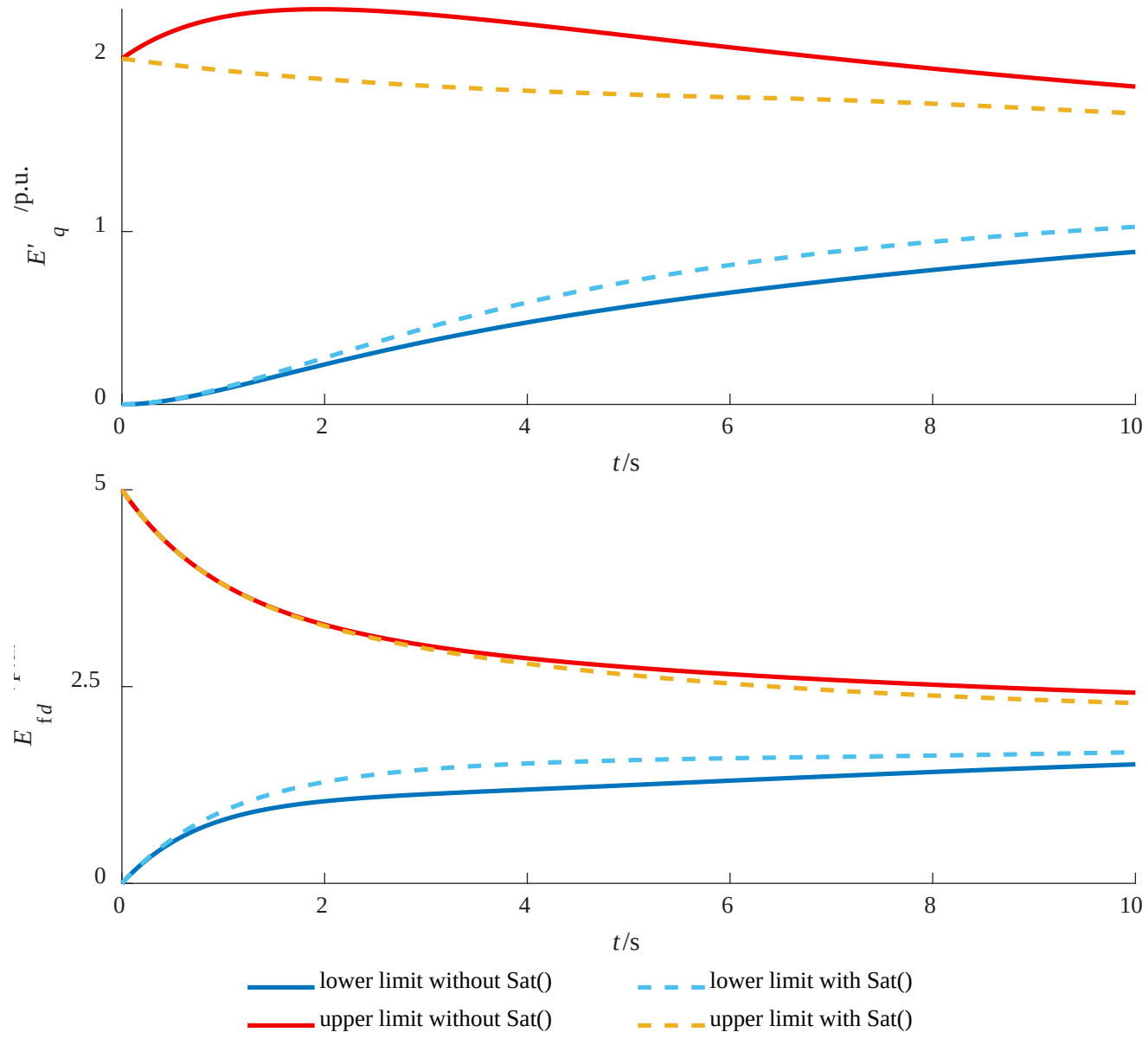

Fig. 5. Time-domain result under different Sat($\boldsymbol{E}_{\mathrm{fd}}$).

### C. Asymptotic Stability of the Subsystem

According to the property of the system Jacobian matrix, the voltage subsystem can also be decomposed into two monotone input-output control systems as shown below:

$$\begin{cases} \boldsymbol{T}'_{d0}\dot{\boldsymbol{E}}'_q = \boldsymbol{K}_{\mathrm{V}}\boldsymbol{E}'_q + \boldsymbol{u}_1 \\ \boldsymbol{u}_1 = \boldsymbol{y}_2 = \boldsymbol{E}_{\mathrm{fd}} \\ \boldsymbol{y}_1 = \boldsymbol{E}'_q \end{cases}, \tag{11}$$

$$\begin{cases} \boldsymbol{T}_{\mathrm{A}}\dot{\boldsymbol{E}}_{\mathrm{fd}} = -\boldsymbol{E}_{\mathrm{fd}} + \boldsymbol{K}_{\mathrm{A}}[\boldsymbol{V}_{\mathrm{ref}} + \boldsymbol{h}'(\boldsymbol{u}_2)] \\ \boldsymbol{h}'(\boldsymbol{u}_2) = -\boldsymbol{h}(-\boldsymbol{u}_2) \\ \boldsymbol{u}_2 = -\boldsymbol{y}_1 = -\boldsymbol{E}'_q \\ \boldsymbol{y}_2 = \boldsymbol{E}_{\mathrm{fd}} \end{cases}. \tag{12}$$

After the above reformulations, since $\partial \boldsymbol{h}/\partial \boldsymbol{E}'_q \geq \boldsymbol{0}$, it appears that $\boldsymbol{E}'_q$ subsystem and $\boldsymbol{E}_{\mathrm{fd}}$ subsystem have the so-called static input-state characteristic, i.e., the steady state changes monotonically with respect to the input [30, 31]. With the help of the following theorem, the equilibrium properties of the voltage subsystem can be studied:

**Theorem 2**[30, 31] If a feedforward-feedback system satisfies:

a) The two subsystems have continuous static input-state characteristics $\boldsymbol{k}_{x1}(\bullet)$ and $\boldsymbol{k}_{x2}(\bullet)$ with negative path;

b) The discrete system $\boldsymbol{u}_{k+1} = -\boldsymbol{k}_{x1}(\boldsymbol{k}_{x2}(\boldsymbol{u}_k))$ has an asymptotic equilibrium $\bar{\boldsymbol{u}}$;

then the original system also has an asymptotic equilibrium: $[\boldsymbol{k}_{x1}(\boldsymbol{k}_{x2}(\bar{\boldsymbol{u}})), \boldsymbol{k}_{x2}(\bar{\boldsymbol{u}})]$.

Fig. 6 vividly illustrates the iterative process of such a discrete system. Since the solutions of a feedforward-feedback system and the associated discrete system are bijective, it provides a way to establish the stability of a feedforward-feedback system by analyzing its simpler discrete system.

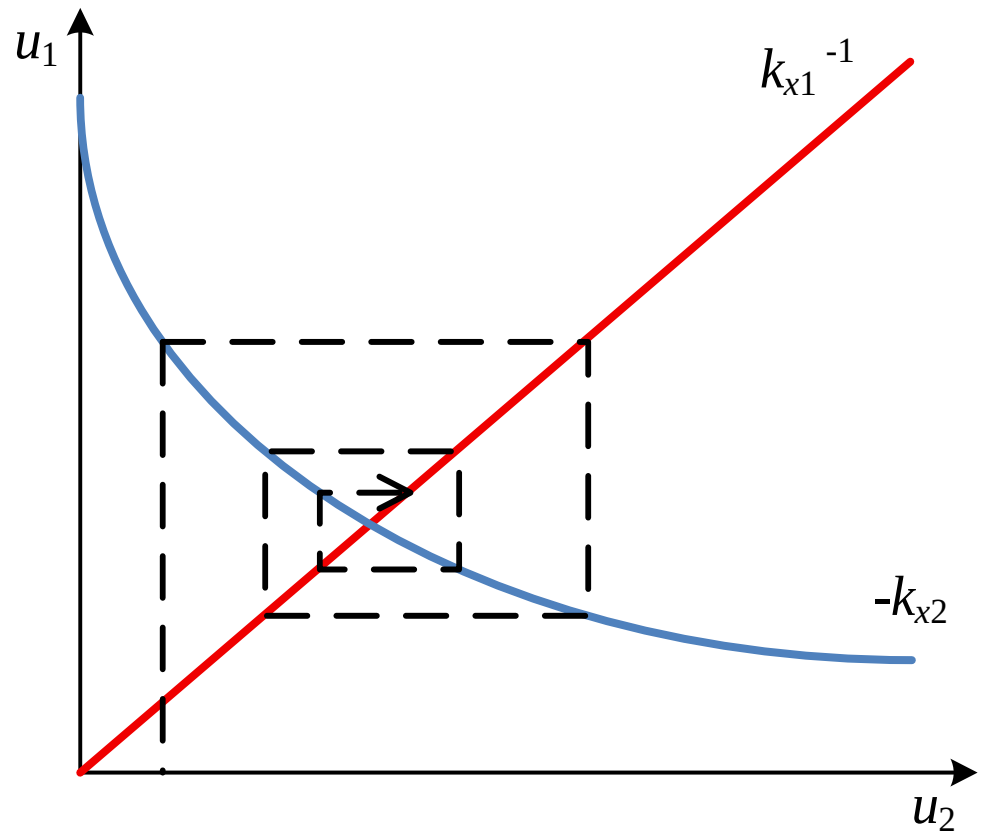

Fig. 6. Schematic diagram of discrete iteration process.

For subsystems (11) and (12), their static characteristics and the associated discrete system can be easily obtained as follows:

$$\begin{cases} \boldsymbol{k}_{x1}(\boldsymbol{u}) = -\boldsymbol{K}_{\mathrm{V}}^{-1}\boldsymbol{u} \\ \boldsymbol{k}_{x2}(\boldsymbol{u}) = \boldsymbol{K}_{\mathrm{A}}\boldsymbol{h}'(\boldsymbol{u}) + \boldsymbol{K}_{\mathrm{A}}\boldsymbol{V}_{\mathrm{ref}} \end{cases}$$

$$\boldsymbol{u}_{k+1} = \boldsymbol{K}_{\mathrm{V}}^{-1}\boldsymbol{K}_{\mathrm{A}}\boldsymbol{h}'(\boldsymbol{u}_k) + \boldsymbol{K}_{\mathrm{V}}^{-1}\boldsymbol{K}_{\mathrm{A}}\boldsymbol{V}_{\mathrm{ref}}$$

To look into the behavior of the above discrete system, the expression of the Jacobian matrix $\boldsymbol{J}_{\mathrm{D}}$ is derived as follows:

$$\boldsymbol{J}_{\mathrm{D}} = \boldsymbol{K}_{\mathrm{V}}^{-1}\boldsymbol{K}_{\mathrm{A}}\frac{\partial \boldsymbol{h}'(\boldsymbol{u})}{\partial \boldsymbol{u}} = \boldsymbol{K}_{\mathrm{V}}^{-1}\boldsymbol{K}_{\mathrm{A}}\frac{\partial \boldsymbol{h}(\boldsymbol{u})}{\partial \boldsymbol{u}},$$

$$J_{\mathrm{D},ij} = \sum_{k=1}^{n} K_{\mathrm{V},ik}^{-1}K_{\mathrm{A},kk}\frac{\partial h_k}{\partial u_j} = \sum_{k=1}^{n} K_{\mathrm{V},ik}^{-1}K_{\mathrm{A},kk}\frac{\sum_{l=1}^{n}\left(C_{k,jl}+C_{k,lj}\right)E'_{q,l}}{2\sqrt{\sum_{p=1}^{n}\sum_{r=1}^{n}C_{k,pr}E'_{q,p}E'_{q,r}}}.$$

Since $-\boldsymbol{K}_{\mathrm{V}}$ is typically an M-matrix, the inverse matrix $\boldsymbol{K}_{\mathrm{V}}^{-1}$ must be negative, this indicates that $\boldsymbol{J}_{\mathrm{D}}$ is a negative matrix. To understand the properties of this negative matrix, we invoke the following result [32]:

1) If $\boldsymbol{A}, \boldsymbol{B}, \boldsymbol{C}, \boldsymbol{D} > \boldsymbol{0}$, $\boldsymbol{A} - \boldsymbol{B} > \boldsymbol{0}$ and $\boldsymbol{C} - \boldsymbol{D} > \boldsymbol{0}$, then $\boldsymbol{AC} > \boldsymbol{BD}$;

2) If $\boldsymbol{A}, \boldsymbol{B} > \boldsymbol{0}$ and $\boldsymbol{A} - \boldsymbol{B} > \boldsymbol{0}$, then $\rho(\boldsymbol{A}) > \rho(\boldsymbol{B})$.

As a result, the spectral radius $\rho(\boldsymbol{J}_{\mathrm{D}})$ is monotone with respect to the entries of $\boldsymbol{J}_{\mathrm{D}}$. In particular, it is concluded that the spectral radius $\rho(\boldsymbol{J}_{\mathrm{D}})$ is a monotone function of coefficients $K_{\mathrm{A}i}$, $i = 1, \ldots n$.

It is well known that an iterative process starting from an open neighborhood of the equilibrium converges necessarily if the spectral radius of the corresponding Jacobian matrix is less than one [33, Section 10.1.3 Ostrowski Theorem]. An implication is that the voltage subsystem must be asymptotically stable before $K_{\mathrm{A}}$ reaches to a critical value to break the voltage stability. In this sense, the spectral radius $\rho(\boldsymbol{J}_{\mathrm{D}})$ can be viewed as a measure of voltage stability. A contraction-type sufficient condition for discrete system stability is also available in [33], we leave this as a topic for future research.

## V. Stability of Closed-loop System

As Section III and Section IV describe, the natures of the feedforward and feedback subsystems are rather different: the feedback subsystem enjoys an elegant cooperativity property and hence exhibits simpler dynamic behavior, while the feedforward subsystem does not. In other words, in voltage dynamics, all synchronous machines behave in a cooperative fashion, while in rotor dynamics, the behaviors of synchronous machines can vary radically.

Having understood the natures of the feedforward and feedback subsystems, one can not stop asking, what stability properties do the closed-loop system bear? In this section we provide an answer to this question, based on the so-called Input-State Stability (ISS) theory [34]. The idea of ISS theory is to determine the closed-loop stability of a control system through the open-loop characteristics of the subsystems. When applying ISS theory to examine the stability of a closed-loop system, one proceeds in two steps: in the first step, one establishes the Local Input-State Stability (LISS) of the subsystems and computes their gains; in the second step, the stability of the closed-loop system is examined using a small gain argument.

### A. LISS of subsystems

To conclude the stability of the closed-loop system, it is required that the feedforward and feedback subsystem be LISS. For a formal definition of LISS, the readers are referred to [35]. The following theorem guarantees that the subsystems have LISS property.

**Theorem 3**[36] For a nonlinear system $\dot{\boldsymbol{x}} = \boldsymbol{f}(\boldsymbol{x}, \boldsymbol{u})$, if the function $\boldsymbol{f}$ is continuously differentiable and the related zero-input system $\dot{\boldsymbol{x}} = \boldsymbol{f}(\boldsymbol{x}, \boldsymbol{0})$ is asymptotically stable, then the system $\dot{\boldsymbol{x}} = \boldsymbol{f}(\boldsymbol{x}, \boldsymbol{u})$ is locally input-state stable.

In Section III and Section IV, the stability properties of subsystems have been studied. In what follows, the computation of the subsystem LISS parameters is unfolded. First, the input/output signals of the subsystems are identified:

$$\begin{cases} \tilde{\boldsymbol{\delta}} = \boldsymbol{\delta}' - \boldsymbol{\delta}_{\mathrm{e}} \\ \tilde{\boldsymbol{E}}'_q = \boldsymbol{E}'_q - \boldsymbol{E}'_{q\mathrm{e}} \end{cases},$$

where $\boldsymbol{\delta}'$ is the relative rotor angle based on the inertial center coordinate; $\boldsymbol{\delta}_{\mathrm{e}}$ and $\boldsymbol{E}'_{q\mathrm{e}}$ represent the equilibrium of the rotor subsystem and the voltage subsystem, respectively.

For simplicity and without loss of generality, it is assumed that the LISS property of each subsystem consists of an exponential decay part with respect to the initial value and a linear gain part with respect to the input. The mathematical form is as follows:

$$\begin{cases} \|\boldsymbol{x}_i\|_2 \le \beta_i \|\xi_i\|_2 e^{-\lambda_i t} + \gamma_i \|\boldsymbol{u}_i\|_\infty \\ \forall t \ge 0, \|\xi_i\|_2 \le v_i, \|\boldsymbol{u}_i\|_\infty \le w_i \end{cases}$$

where $\|\cdot\|_2$ denotes the Euclidean norm and $\|\cdot\|_\infty$ denotes the infinite norm; $\boldsymbol{x}$ is the state quantity, $\boldsymbol{\xi}$ is the initial state value and $\boldsymbol{u}$ is the input; $t$ indicates the time, $v$ is the state constraint and $w$ is the input constraint; $\beta$ is the initial value gain coefficient, $\lambda$ is the decay coefficient and $\gamma$ is the input gain coefficient; the subscript $i$ indicates the $i$-th subsystem.

The determination of LISS parameters may be analytically deduced for low-dimensional systems. But for high-dimensional systems, it is more convenient to use simulation data to estimate parameters. The estimation process is summarized as follows [37]:

1) Given the initial state $\boldsymbol{\xi}$ and the input $\boldsymbol{u}$, use a numerical integration method to calculate the time-domain response $\boldsymbol{x}(t)$.

2) Reduce the initial state $\boldsymbol{\xi}$ or the input $\boldsymbol{u}$, until the subsystem returns to stability. The critical values are recorded as the state constraint $v$ and the input constraint $w$.

3) If the subsystem remains stable, record the steady state as $\boldsymbol{x}_{\mathrm{e}}$. Since the initial value response will decay to 0 as $t\to\infty$, the input gain coefficient $\gamma$ can be calculated as follows:

$$\gamma = \lim_{t\to\infty} \|\boldsymbol{x}_e\|_2 \big/ \|\boldsymbol{u}\|_\infty . \tag{13}$$

4) Vary the input $\boldsymbol{u}$ to obtain a series of input gain coefficients. The maximum value among them is taken as the estimation result of the input gain coefficient.

5) Record the maximum value of the state response as $\boldsymbol{x}_{\max}$ and the corresponding time as $t_{\max}$. Likewise, record the time close to the steady state as $t_{\mathrm{end}}$ and the corresponding state as $\boldsymbol{x}_{\mathrm{end}}$. Then the decay coefficient $\lambda$ and the initial value gain coefficient $\beta$ can be calculated as:

$$\lambda = \frac{1}{t_{\max} - t_{\mathrm{end}}} \ln\left( \frac{\|\boldsymbol{x}_{\max}\|_2 - \gamma \|u\|_\infty}{\|\boldsymbol{x}_{\mathrm{end}}\|_2 - \gamma \|u\|_\infty} \right),$$

$$\beta = \frac{\|\boldsymbol{x}_{\max}\|_2 - \gamma \|\boldsymbol{u}\|_\infty}{\|\xi\|_2} e^{-\lambda t_{\mathrm{end}}} .$$

6) Vary the input $\boldsymbol{u}$ and the initial value $\boldsymbol{\xi}$ to obtain a series of $\lambda$ and $\beta$. The smallest $\lambda$ and the largest $\beta$ are taken as the final estimation result.

The above procedure yields the desired parameters $w$, $v$, $\gamma$, $\lambda$ and $\beta$, which will be used in the next step to establish the stability of the closed-loop system.

Here we briefly discuss the impacts of subsystem parameters on LISS properties. In the rotor subsystem, with the increase of the inertia constant $H$, the system synchronization ability is enhanced. As a result, the state constraint $v_1$ and the input constraint $w_1$ increase, but the decay coefficient $\lambda_1$ decreases. However, it has no impact on the input gain coefficient $\gamma_1$ because the equilibrium point has not changed.

In the voltage subsystem, $T_{\mathrm{A}}$ is much smaller than $T'_{d0}$, so there will be a peak in the time-domain response of the excitation voltage. When $T_{\mathrm{A}}$ is further reduced, the regulating speed becomes faster and the stability margin becomes worse. As a result, the state constraint $v_2$ and the input constraint $w_2$ decrease, but the decay coefficient $\lambda_2$ increases. However, it has no impact on the input gain coefficient $\gamma_2$ because the equilibrium point has not changed. In addition, an increase in $K_{\mathrm{A}}$ amplifies the tracking bias, so the decay coefficient $\lambda_2$ will increases, while the state constraint $v_2$ and the input constraint $w_2$ decrease.

*B. Stability criterion of closed-loop system*

In this section, with the help of the celebrated Small-Gain-Theorem, the stability of the closed-loop system is established. As can be seen from Fig. 1 the output of any one subsystem is directly connected to the input of the other subsystem, forming a closed-loop system with no external input. Let us define a matrix $\boldsymbol{Z}$ to describe such an input-output connection relationship:

$$\boldsymbol{Z} = \begin{bmatrix} 0 & 1 \\ 1 & 0 \end{bmatrix}.$$

And let matrix $\boldsymbol{G}$ represent the loop gains of the subsystems:

$$\boldsymbol{G}_L = \boldsymbol{\gamma}\boldsymbol{Z} = \begin{bmatrix} 0 & \gamma_1 \\ \gamma_2 & 0 \end{bmatrix},$$

where $\boldsymbol{\gamma}$ denotes the diagonal matrix formed by the input gains of subsystems; subscripts 1 and 2 indicate the feedforward and the feedback subsystem, respectively.

The theorem below establishes the closed-loop stability of the feedforward-feedback system consisting of two LISS subsystems.

**Theorem 4** [38] A closed-loop system with an initial value $\boldsymbol{\xi}$ and no external input is asymptotically stable when:

a) The subsystems are LISS;

b) Small gain condition:

$$\rho(\boldsymbol{G}_L) = \sqrt{\lambda_1 \lambda_2} < 1,$$

where $\rho(\cdot)$ represents the spectral radius of a matrix;

c) Boundary constraint:

$$\boldsymbol{Z}(\mathbf{I}_2 - \boldsymbol{G}_L)^{-1} \boldsymbol{\beta} \left[ \|\xi_1\|_2, \|\xi_2\|_2 \right]^{\mathrm{T}} \le \left[ w_1, w_2 \right]^{\mathrm{T}}, \tag{14}$$

where $[\|\xi_1\|_2, \|\xi_2\|_2]^{\mathrm{T}} \le [v_1, v_2]^{\mathrm{T}}$ and $\boldsymbol{\beta} = [\beta_1, 0; 0, \beta_2]$.

The LISS parameter estimation results and the closed-loop stability conditions are highly correlated with the specific decomposition-aggregation strategy. In general, the closed-loop system enjoys higher stability margin if each individual subsystem exhibits lower gain.

We end this section by noting that, the suggested decomposition-aggregation strategy allows one to look into the contribution of each individual subsystem. This topic, which is of apparent interest, will be pursued in future research.

## VI. Case Studies

In this section the properties of the voltage subsystem are illustrated using several examples, then the detailed LISS estimation results of the rotor subsystem and the voltage subsystem are shown.

### A. *Properties of voltage subsystem*

In the 3-machine test system, a 3-phase short-circuit fault occurs at one end of Line 5_7 at $t_1$ = 0s, and then it is cleared by tripping the faulted line at $t_2$ = 0.05s. The partial derivative results are plotted in Fig. 7, showing that all entries of the Jacobian matrix remain sign stable.

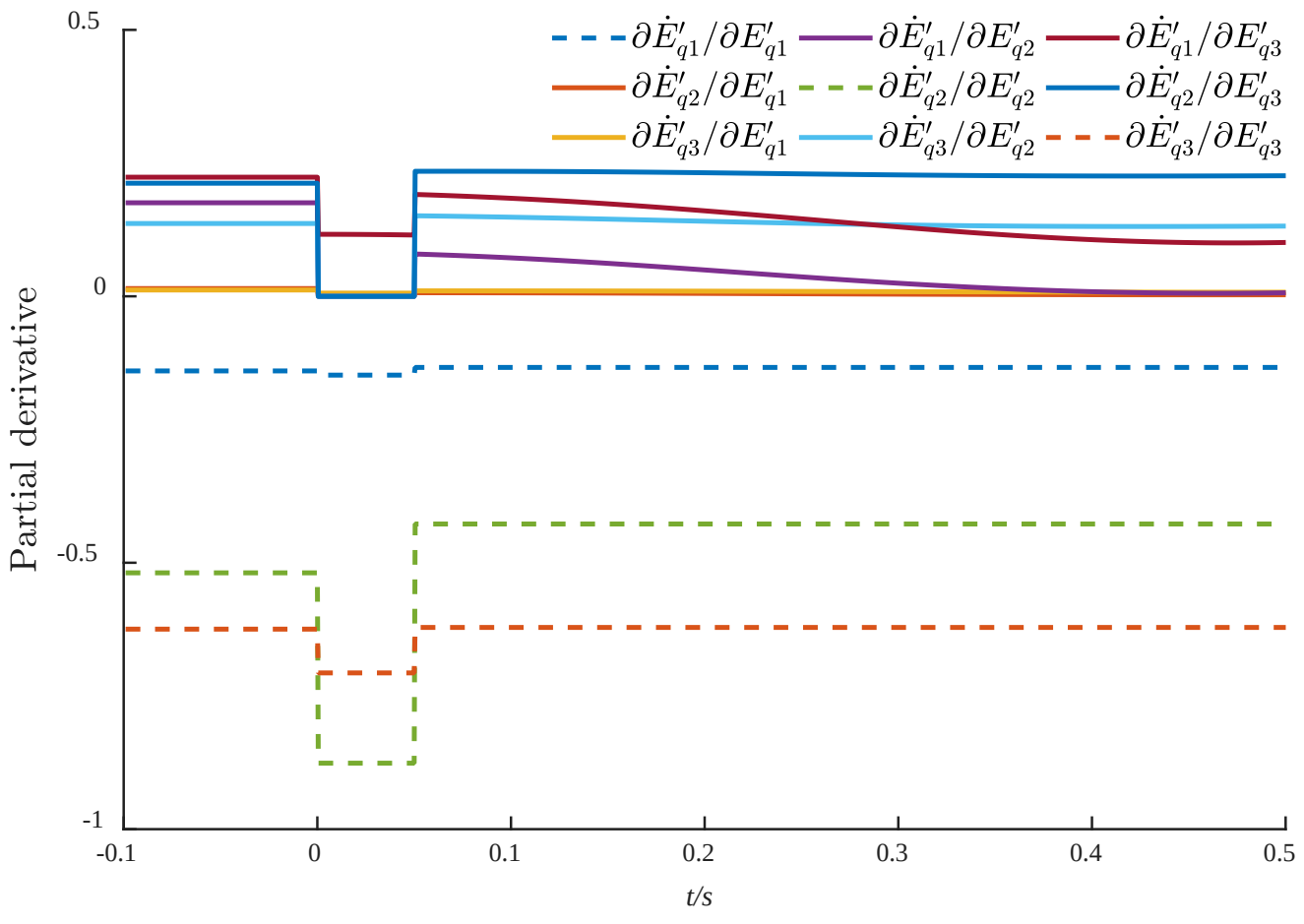

(a) $\partial\dot{\boldsymbol{E}}'_q \,/\, \partial\boldsymbol{E}'^{\mathrm{T}}_q$

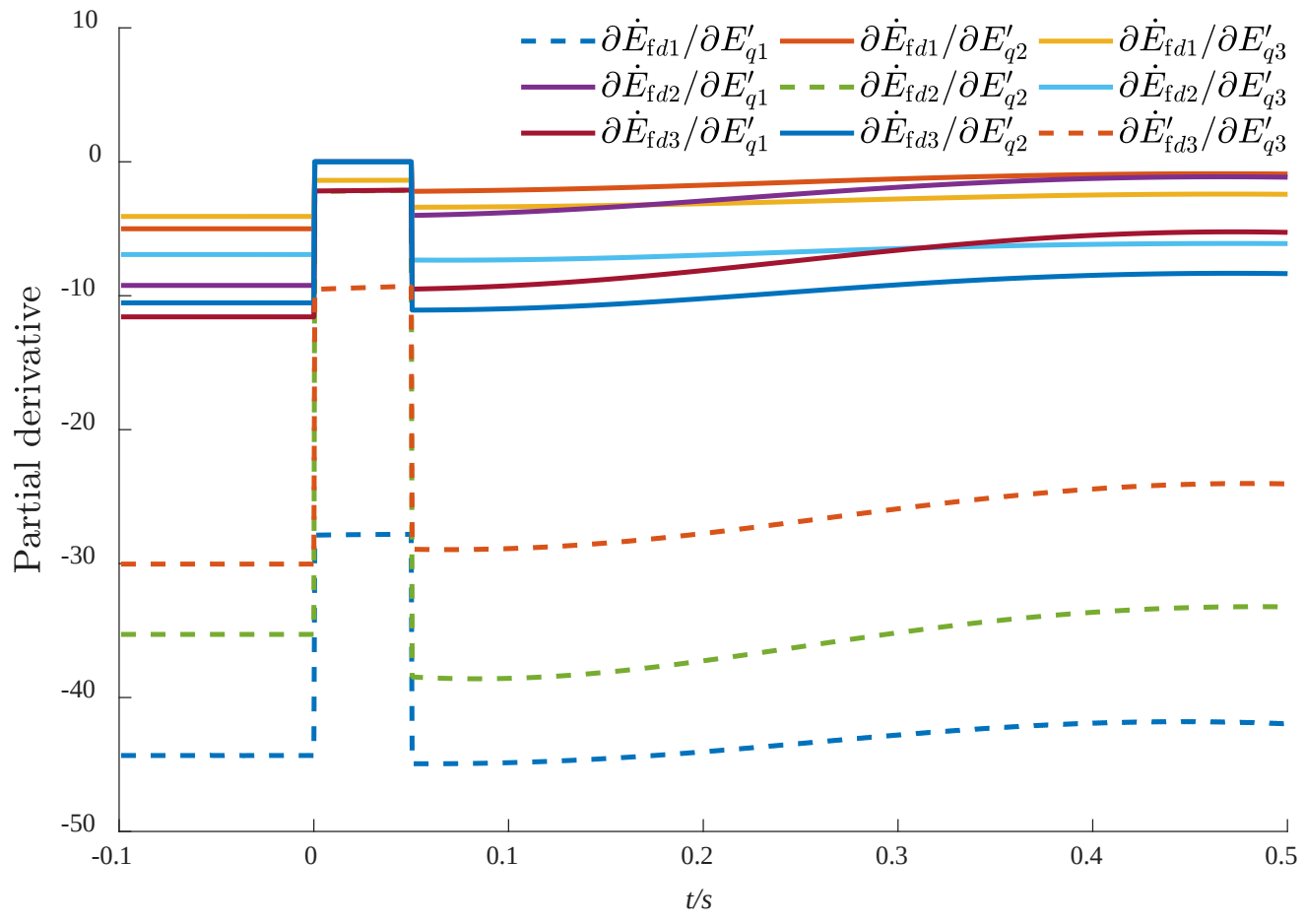

(b) $\partial\dot{\boldsymbol{E}}_{\mathrm{fd}} \,/\, \partial\boldsymbol{E}'^{\mathrm{T}}_q$

Fig. 7. Time-domain partial derivatives of non-fixed entries.

In a 50-machine system [39], a 3-phase short-circuit fault occurs in Bus_120, and then it is cleared after 0.05s. The partial derivatives are also calculated to confirm the sign pattern of the system Jacobian matrix. The sign result is reported in Fig.8, similar to (10), indicating that the Jacobian matrix of the power system is sign-definite.

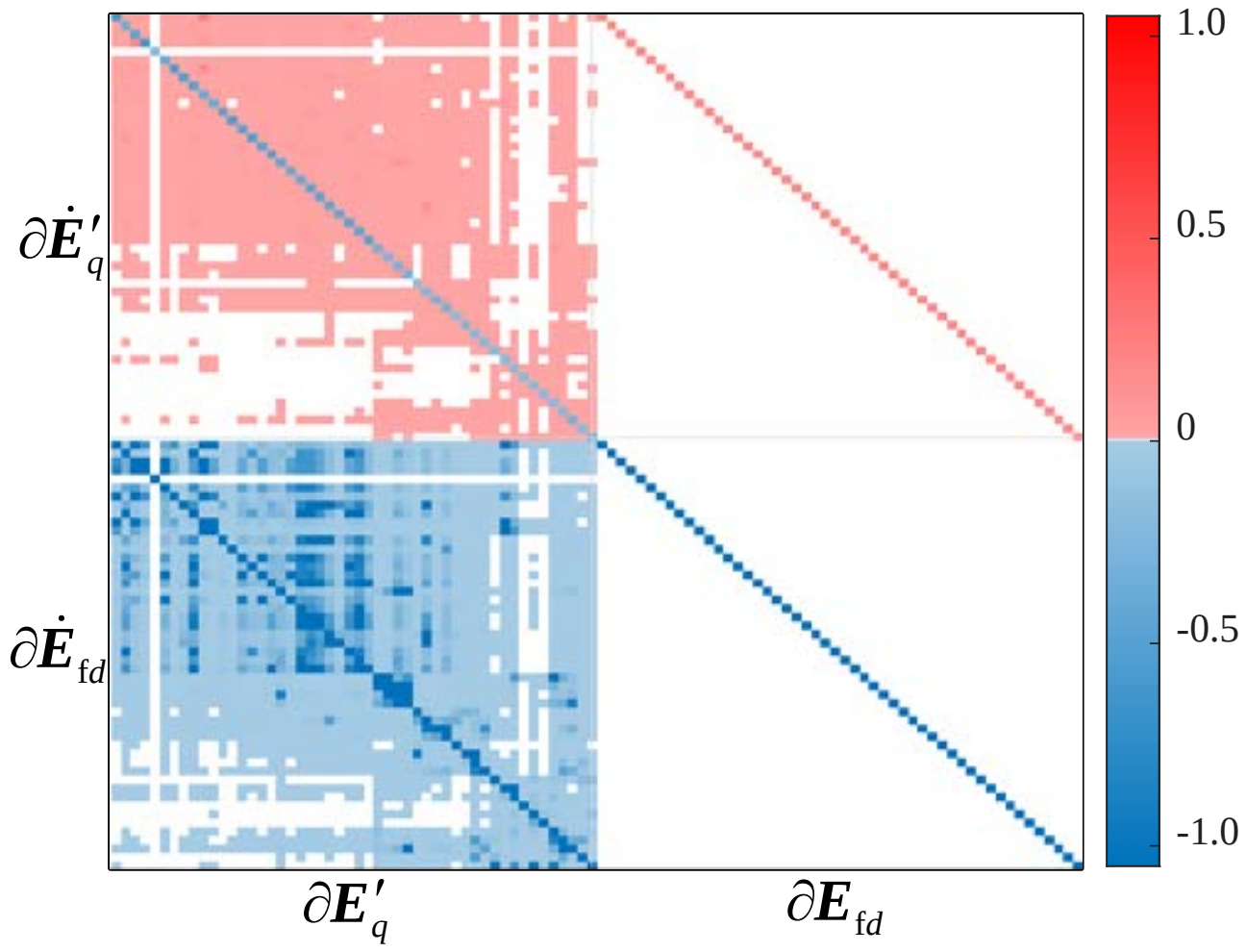

Fig. 8. Sign pattern of the 50-machine system.

Based on the above-mentioned sign patterns existing in the original Jacobian matrix, an augmented system can be easily constructed. Its trajectories describe the upper and lower limits of the time-domain voltage response, which can be used to analyze the effect of exciter parameters. For example, when exciter time constant $T_A$ takes the value of 1.0/0.5/0.1 respectively, the time-domain responses will change accordingly, as shown in Fig. 9. It can be seen that the variation of $T_A$ has no essential impact on the ultimate convergence property of the trajectory, and the convergence rate improves as $T_A$ decrease, also the difference between the trajectory bound and the actual trajectory reduces.

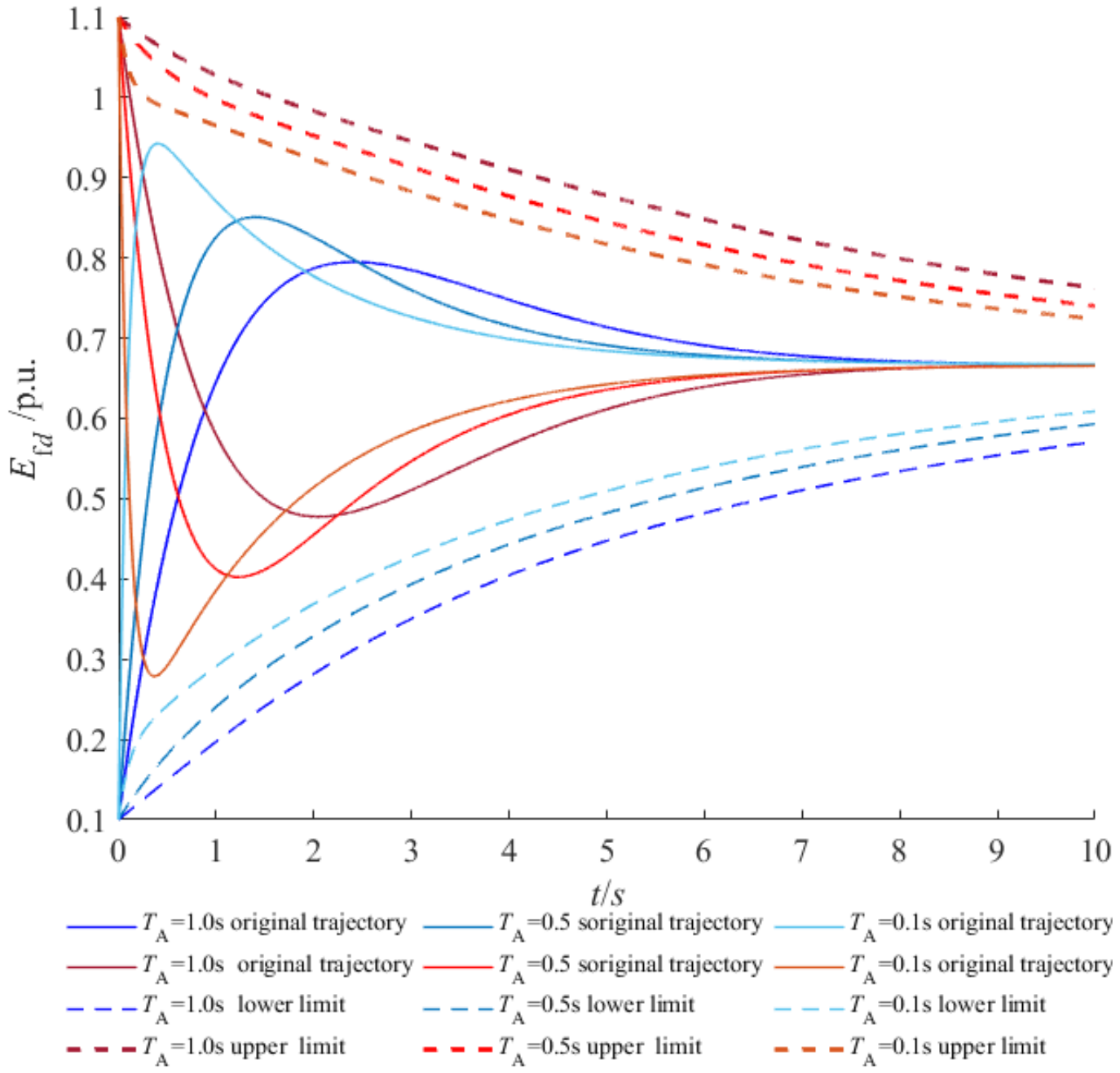

Fig. 9. Time-domain result of $E_{fdvalues}$ under different $T_A$.

### B. *Stability results of closed-loop system*

We start from an idealistic single machine infinite bus (SMIB) test system. It can be decomposed into two subsystems, as described in Section II. The time-domain response $\boldsymbol{x}(t)$ and the steady-state value $\boldsymbol{x}_e$ corresponding to different inputs are calculated by numerical integration method for each subsystem. In step 2), the input constraint and the state constraint of two subsystems are determined as $w_1$ = 0.48, $v_1$ = 0.40, $w_2 = \pi$ and $v_2$ = 1.47. Referring to (13), the input gain coefficients $\gamma_1$ and $\gamma_2$ under different inputs are obtained as shown in Fig. 10.

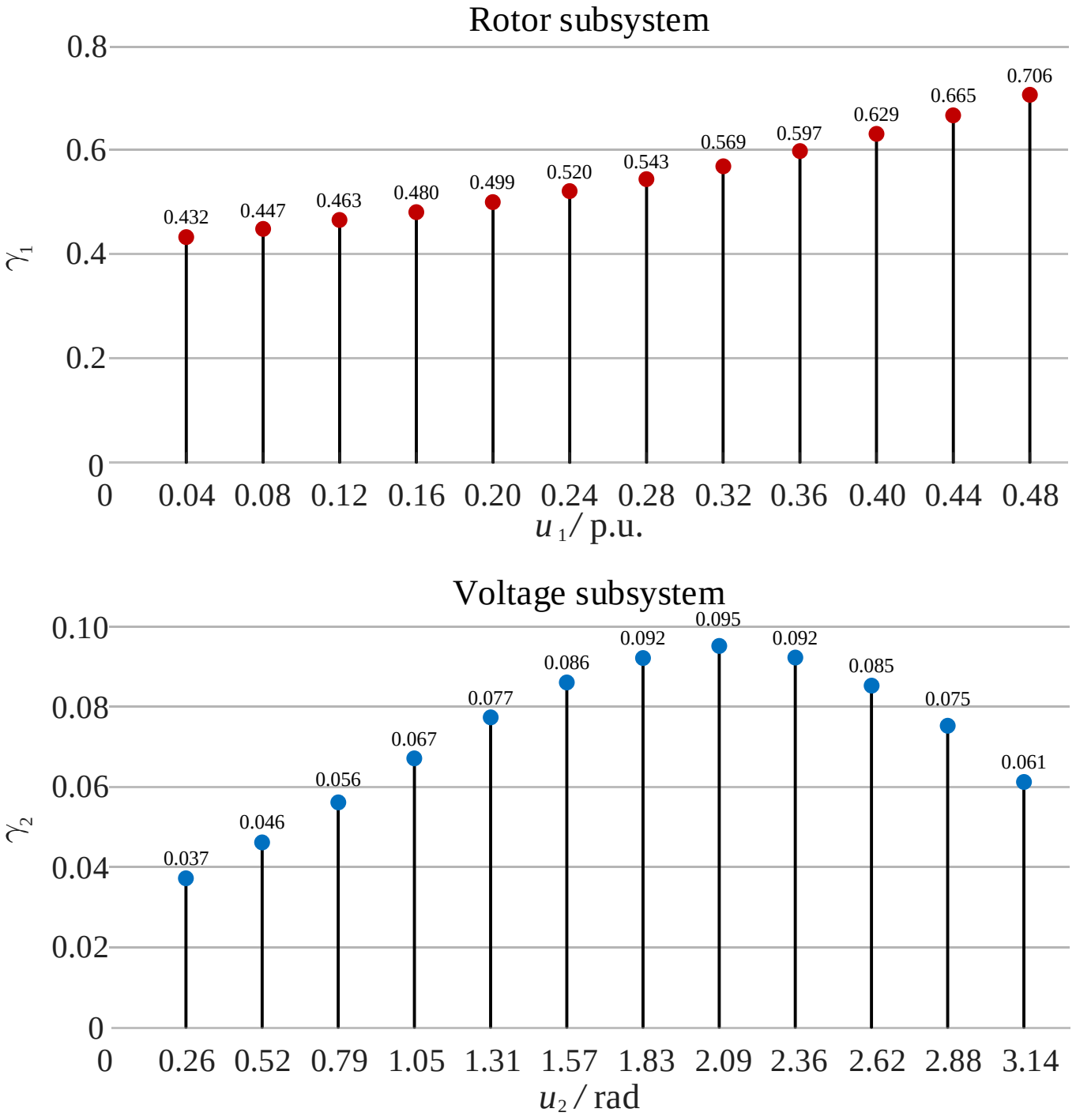

Fig. 10. Input gain coefficients in SMIB system.

Take the maximum as estimation results: $\gamma_1$ = 0.706 and $\gamma_2$ = 0.095. Both are less than 1, reflecting the weak coupling between them. By varying the initial value $\xi$, a series of time-domain responses $\boldsymbol{x}(t)$ are calculated to determine the decay coefficient $\lambda$ and the initial value gain coefficient $\beta$ in step 5). The results are shown in Table 1.

Table 1. Initial gain of SMIB system.

| $\|\|\xi_1\|\|_2$ | $\lambda_1$ | $\beta_1$ |
|---|---|---|
| 0.10 | 0.064 | 1.007 |
| 0.20 | 0.063 | 0.992 |
| 0.40 | 0.062 | 1.000 |
| $\|\|\xi_2\|\|_2$ | $\lambda_2$ | $\beta_2$ |
| 0.25 | 1.124 | 0.992 |
| 0.50 | 1.128 | 0.996 |
| 1.00 | 1.131 | 0.998 |

It is worth noting that the $\gamma$ used to determine $\lambda$ and $\beta$ is the maximum value for different cases, which introduces conservatism. Therefore, the $\beta$ with input will be smaller than the result without input. Taking the maximum and minimum values as estimation results, one obtains $\lambda_1$ = 0.0623, $\beta_1$ = 1.007, $\lambda_2$ = 1.124 and $\beta_2$ = 0.998. Therefore, the spectral radius of $\boldsymbol{G}$ is determined as follows:

$$\rho(\boldsymbol{G}_L) = \sqrt{\gamma_1\gamma_2} = 0.252 < 1 .$$

The result indicates that the small gain condition is satisfied. Substituting $\|\xi_1\|_2$ = 0.40 and $\|\xi_2\|_2$ = 0.41 into the left side of (14), one obtains:

$$\boldsymbol{Z}(\mathbf{I}_2 - \boldsymbol{G}_L)^{-1}\boldsymbol{\beta}\begin{bmatrix}0.40\\0.41\end{bmatrix} = \begin{bmatrix}0.479\\0.741\end{bmatrix}. \qquad (15)$$

The above result does not exceed the boundary constraint $[w_1, w_2]^{\mathrm{T}} = [0.48, \pi]^{\mathrm{T}}$, so according to Theorem 4, the SMIB system under the given initial value is asymptotically stable. The related time-domain simulations help to confirm the asymptotic stability. Besides, using the numerical integration method, the actual critical $\|\xi_2\|_2$ is determined to be 0.67 corresponding to $|\xi_1\|_2$ = 0.40. It is slightly larger than the estimation value in (15).

Furthermore, the regulator gain coefficient $K_A$ has a key impact on the LISS properties of the voltage subsystem, which in turn affects the closed-loop stability of power systems. Fig.11 shows that the input gain coefficient $\gamma_2$ increases as $K_A$ increases.

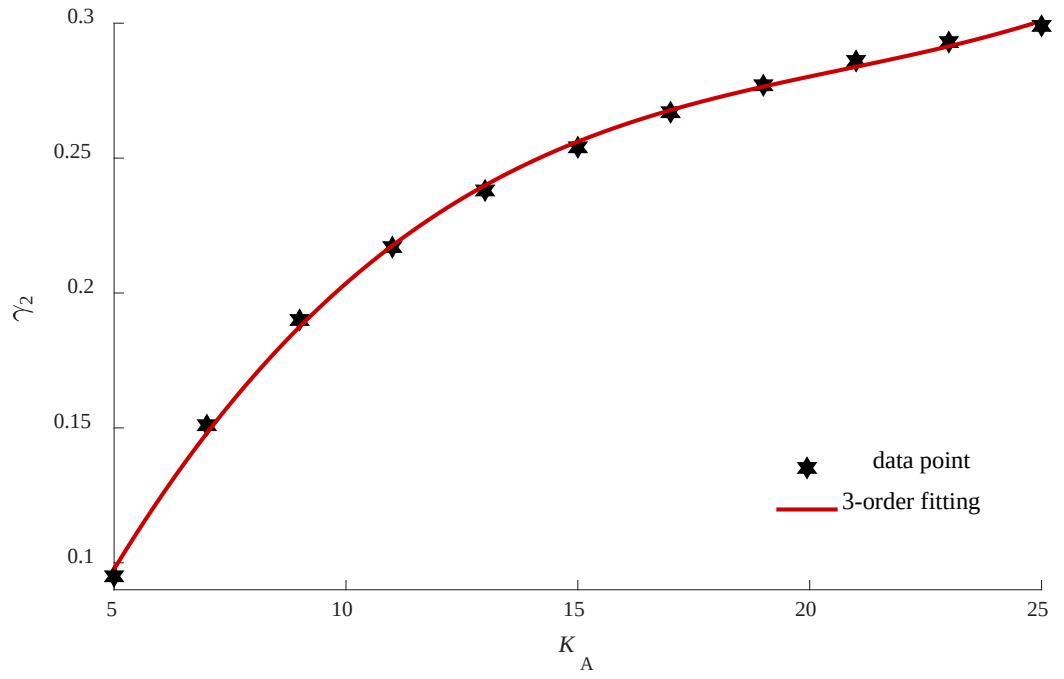

Fig. 11. Input gain coefficient $\gamma_2$ changes smoothly with $K_A$.

Applying the third-order polynomial fitting to establish the input-output relationship between $\gamma_2$ and $K_A$, one obtains:

$$\gamma_2 = 0.00003K_A^3 - 0.0021K_A^2 + 0.0463K_A - 0.0866 .$$

Referring to the expression, $\gamma_2$ = 0.2445 corresponds to $K_A$ = 13.5. In this case, if given $\|\xi_1\|_2$ = 0.40, $\|\xi_2\|_2$ = 0.30, then

$$\rho(\boldsymbol{G}_L) = \sqrt{\gamma_1\gamma_2} = 0.415 ,$$

$$\boldsymbol{Z}(\mathbf{I}_2 - \boldsymbol{G}_L)^{-1}\boldsymbol{\beta}\begin{bmatrix}0.40\\0.30\end{bmatrix} = \begin{bmatrix}0.480\\0.742\end{bmatrix}.$$

The discussion above means that the upper limit of $K_A$ for stability can be quickly estimated through algebraic operations. Using the numerical integration method, the exact upper limit of $K_A$ is determined to be 17.5. It is slightly larger than the estimation value based on the proposed algebraic operations.

Next, we report the test results of the 3-machine test system. The input constraint and the state constraint of two subsystems are determined as $w_1$ = 0.42, $v_1$ = 0.38, $w_2 = \pi$ and $v_2$ = 1.80. Fig. 12 shows the input gain coefficients obtained under different inputs. It can be seen that the magnitude and trend of the input gain coefficients in the 3-machine system are similar to the results in SMIB system. Taking the maximum value as estimation results, one obtains $\gamma_1$ = 0.92 and $\gamma_2$ = 0.135.

In addition, the results of initial gain are shown in Table 2. Taking the maximum and minimum values as estimation results, one obtains $\lambda_1$ = 0.0381, $\beta_1$ = 1.1714, $\lambda_2$ = 0.7456 and $\beta_2$ = 1.000.

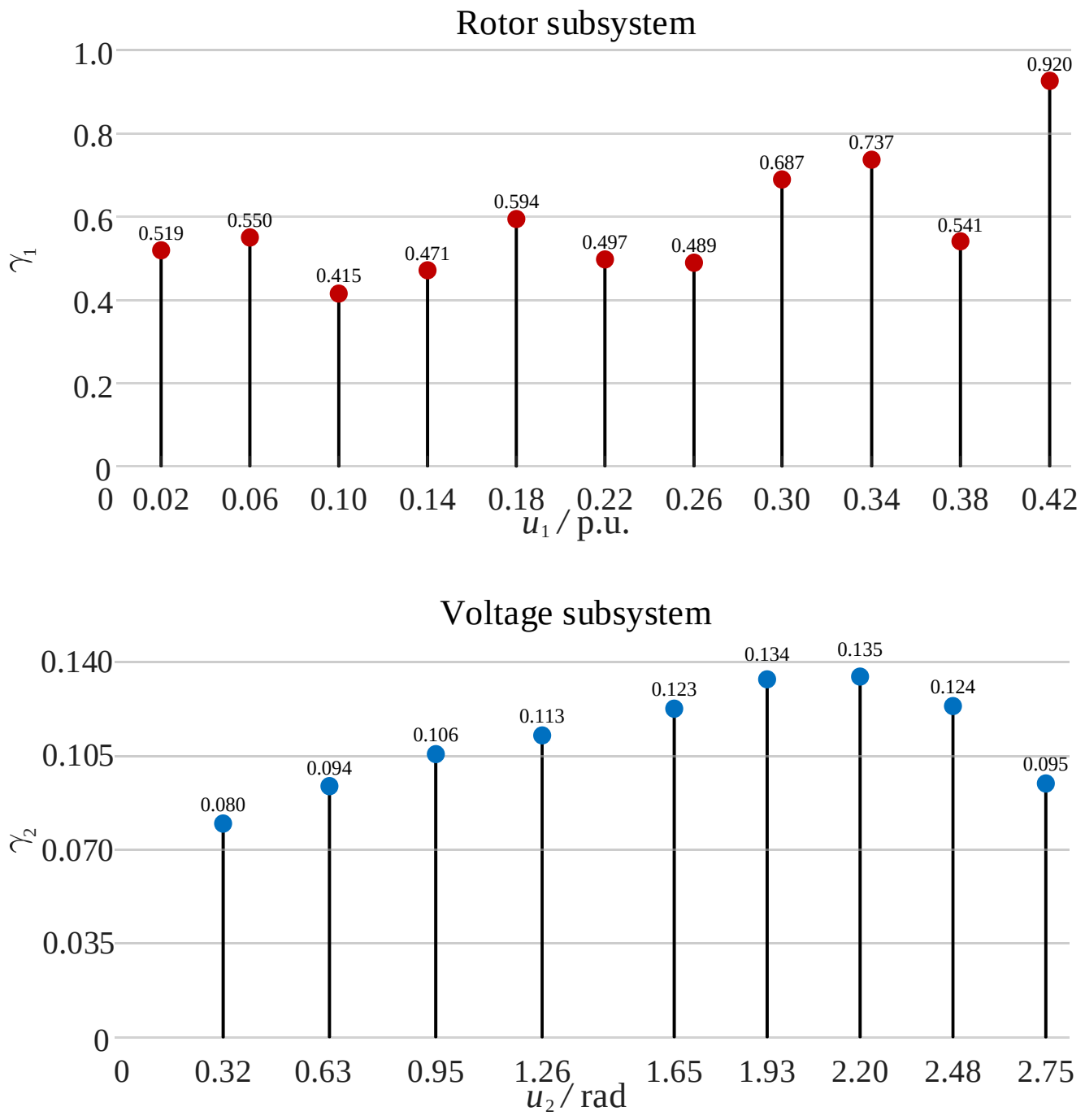

Fig. 12. Input gain coefficients in 3-machine system.

Table 2. Initial gain of the 3-machine system.

| $\|\xi_1\|_2$ | $\lambda_1$ | $\beta_1$ |
|---|---|---|
| 0.085 | 0.0381 | 1.1618 |
| 0.256 | 0.0392 | 1.1673 |
| 0.375 | 0.0385 | 1.1714 |
| $\|\xi_2\|_2$ | $\lambda_2$ | $\beta_2$ |
| 0.25 | 0.7929 | 1.0000 |
| 0.50 | 0.7669 | 1.0000 |
| 1.00 | 0.7456 | 1.0000 |

Therefore, the spectral radius of $\boldsymbol{G}_L$ is determined as follows:

$$\rho(\boldsymbol{G}_L) = \sqrt{\gamma_1\gamma_2} = \sqrt{0.92\times 0.135} = 0.352 < 1 .$$

The result indicates that the small gain condition is satisfied. Substituting $\|\xi_1\|_2 = 0.35$ and $\|\xi_2\|_2 = 0.30$ into the left side of (14), one obtains:

$$\boldsymbol{Z}(\mathbf{I}_2 - \boldsymbol{G}_L)^{-1}\boldsymbol{\beta}\begin{bmatrix}0.35\\0.31\end{bmatrix} = \begin{bmatrix}0.4172\\0.7938\end{bmatrix}. \tag{16}$$

The above result does not exceed the boundary constraint $[w_1, w_2]^{\mathrm{T}} = [0.42, \pi]^{\mathrm{T}}$, so according to Theorem 4, the 3-machine system under the given initial value is asymptotically stable. The asymptotic stability is also verified by time-domain simulations. Besides, the actual critical $\|\xi_2\|_2$ is determined to be 0.57 corresponding to $|\xi_1\|_2 = 0.35$, which is slightly larger than the estimation value in (16).

For the analysis of the initial value disturbance and the limit of control parameter, the estimation results are conservative to a certain extent. However, the small gain condition and the boundary constraint in Local Small-Gain-Theorem provide a concise sufficient criterion in algebraic form for judging the closed-loop stability, which helps reveal the interaction between rotor dynamics and voltage dynamics in power systems.

## VII. Conclusion

The structural properties of a power system dynamic model are studied using a decomposition-aggregation approach. It is demonstrated that the subsystems bear distinct natures, and the system model ought to be decomposed based on the natures of subsystems. This new decomposition-aggregation strategy helps reduce the complexity of stability analysis and provides many insights into power system dynamics which is becoming increasingly complex in recent years.

Many questions remain to be addressed, for example, what are the roles of power system stabilizers? In which subsystem governor and prime-mover should reside [40]? In this regard, the introduced results open up a new avenue of power system dynamics research, endorsed by the theory of interconnected nonlinear systems.

## VIII. Appendix – Extension to Realistic dynamic Models

Taking into account the transient saliency of rotors, the injection current $\boldsymbol{I}_{\mathrm{G}}$ will couple the conjugate of generator terminal voltage ($\bar{\boldsymbol{V}}_{\mathrm{G}}$), which is rewritten as

$$\boldsymbol{I}_{\mathrm{G}} = -\mathrm{j}\boldsymbol{X}_d'^{-1}e^{\mathrm{j}\boldsymbol{\delta}}\boldsymbol{E}_q' - \boldsymbol{Y}_1\boldsymbol{V}_{\mathrm{G}} + \boldsymbol{Y}_2 e^{\mathrm{j}2\boldsymbol{\delta}}\bar{\boldsymbol{V}}_{\mathrm{G}}, \tag{A.1}$$

where $\boldsymbol{Y}_1 = -\mathrm{j}(\boldsymbol{X}_q'^{-1} + \boldsymbol{X}_d'^{-1})/2$ and $\boldsymbol{Y}_2 = -\mathrm{j}(\boldsymbol{X}_q'^{-1} - \boldsymbol{X}_d'^{-1})/2$ ; $e^{\mathrm{j}\boldsymbol{\delta}}$ and $e^{\mathrm{j}2\boldsymbol{\delta}}$ are the diagonal matrices composed of the exponential function of $\boldsymbol{\delta}$, respectively.

Substituting (A.1) into the algebraic network equation gives

$$\begin{bmatrix}\boldsymbol{Y}_{\mathrm{G}}+\boldsymbol{Y}_1 & \boldsymbol{Y}_{\mathrm{GL}}\\ \boldsymbol{Y}_{\mathrm{LG}} & \boldsymbol{Y}_{\mathrm{L}}\end{bmatrix}\begin{bmatrix}\boldsymbol{V}_{\mathrm{G}}\\ \boldsymbol{V}_{\mathrm{L}}\end{bmatrix} = \begin{bmatrix}-\mathrm{j}\boldsymbol{X}_d'^{-1}e^{\mathrm{j}\boldsymbol{\delta}}\boldsymbol{E}_q' + \boldsymbol{Y}_2 e^{\mathrm{j}2\boldsymbol{\delta}}\bar{\boldsymbol{V}}_{\mathrm{G}}\\ \mathbf{0}\end{bmatrix}.$$

After eliminating load voltage $\boldsymbol{V}_{\mathrm{L}}$, $\boldsymbol{V}_{\mathrm{G}}$ is expressed as:

$$\boldsymbol{V}_{\mathrm{G}} = \underbrace{\boldsymbol{Z}_{\mathrm{G}}\boldsymbol{Y}_2 e^{\mathrm{j}2\boldsymbol{\delta}}}_{\boldsymbol{A}_{\mathrm{G}}}\bar{\boldsymbol{V}}_{\mathrm{G}} \underbrace{-\mathrm{j}\boldsymbol{Z}_{\mathrm{G}}\boldsymbol{X}_d'^{-1}e^{\mathrm{j}\boldsymbol{\delta}}}_{\boldsymbol{B}_{\mathrm{G}}}\boldsymbol{E}_q' = \boldsymbol{A}_{\mathrm{G}}\bar{\boldsymbol{V}}_{\mathrm{G}} + \boldsymbol{B}_{\mathrm{G}}\boldsymbol{E}_q',$$

where $\boldsymbol{Z}_{\mathrm{G}} = (\boldsymbol{Y}_{\mathrm{G}} + \boldsymbol{Y}_1 - \boldsymbol{Y}_{\mathrm{GL}}\,\boldsymbol{Y}_{\mathrm{L}}^{-1}\,\boldsymbol{Y}_{\mathrm{LG}})^{-1}$.

Quite often $\boldsymbol{V}_{\mathrm{G}}$ is solved by using Dommel-Sato iteration [41]. The iteration process is shown below:

$$\begin{cases} \boldsymbol{V}_{\mathrm{G}}^{(1)} = \boldsymbol{A}_{\mathrm{G}}\bar{\boldsymbol{V}}_{\mathrm{G}}^{(0)} + \boldsymbol{B}_{\mathrm{G}}\boldsymbol{E}_q' \\ \boldsymbol{V}_{\mathrm{G}}^{(2)} = \boldsymbol{A}_{\mathrm{G}}\bar{\boldsymbol{V}}_{\mathrm{G}}^{(1)} + \boldsymbol{B}_{\mathrm{G}}\boldsymbol{E}_q' = \boldsymbol{A}_{\mathrm{G}}\bar{\boldsymbol{A}}_{\mathrm{G}}\boldsymbol{V}_{\mathrm{G}}^{(0)} + \boldsymbol{A}_{\mathrm{G}}\bar{\boldsymbol{B}}_{\mathrm{G}}\boldsymbol{E}_q' + \boldsymbol{B}_{\mathrm{G}}\boldsymbol{E}_q' \\ \quad\vdots \\ \boldsymbol{V}_{\mathrm{G}}^{(2n)} = \sum_{i=0}^{n-1}(\boldsymbol{A}_{\mathrm{G}}\bar{\boldsymbol{A}}_{\mathrm{G}})^i(\boldsymbol{B}_{\mathrm{G}} + \boldsymbol{A}_{\mathrm{G}}\bar{\boldsymbol{B}}_{\mathrm{G}})\boldsymbol{E}_q' + (\boldsymbol{A}_{\mathrm{G}}\bar{\boldsymbol{A}}_{\mathrm{G}})^n\boldsymbol{V}_{\mathrm{G}}^{(0)} \end{cases},$$

where the superscript of $\boldsymbol{V}_{\mathrm{G}}$ indicates the number of iterations.

The computational experience points out that the spectral radius of $\boldsymbol{A}_{\mathrm{G}}\bar{\boldsymbol{A}}_{\mathrm{G}}$ is less than 1, so the sum of $(\boldsymbol{A}_{\mathrm{G}}\bar{\boldsymbol{A}}_{\mathrm{G}})^i$ is convergent. The nice consequence of this fact is that $\boldsymbol{V}_{\mathrm{G}}$ has a closed-form expression as follows:

$$\boldsymbol{V}_{\mathrm{G}} = (\mathbf{I}_m - \boldsymbol{A}_{\mathrm{G}}\bar{\boldsymbol{A}}_{\mathrm{G}})^{-1}(\boldsymbol{B}_{\mathrm{G}} + \boldsymbol{A}_{\mathrm{G}}\bar{\boldsymbol{B}}_{\mathrm{G}})\boldsymbol{E}_q' = (\boldsymbol{K}_X + \mathrm{j}\boldsymbol{K}_Y)\boldsymbol{E}_q',$$

where $\boldsymbol{K}_X$ and $\boldsymbol{K}_Y$ are real-valued matrices.

Moreover, $\boldsymbol{I}_d$ has the following expression:

$$\begin{aligned}\boldsymbol{I}_d &= \boldsymbol{X}_d'^{-1}(\boldsymbol{E}_q' - \boldsymbol{V}_q) = \boldsymbol{X}_d'^{-1}(\boldsymbol{E}_q' - \boldsymbol{V}_x \circ \cos\boldsymbol{\delta} - \boldsymbol{V}_y \circ \sin\boldsymbol{\delta}) \\ &= \underbrace{\boldsymbol{X}_d'^{-1}(\mathbf{I}_n - \boldsymbol{K}_X\cos\boldsymbol{\delta} - \boldsymbol{K}_Y\sin\boldsymbol{\delta})}_{\boldsymbol{g}(\boldsymbol{\delta})}\boldsymbol{E}_q' = \boldsymbol{g}(\boldsymbol{\delta})\boldsymbol{E}_q'\end{aligned},$$

where $\boldsymbol{V}_x$ and $\boldsymbol{V}_y$ represent the generator terminal voltage on $x$-axis and $y$-axis, respectively.

From the expressions of $\boldsymbol{I}_d$, it is not difficult to see that, the power system dynamic model can still be represented by a set of pure ordinary differential equations. However, if loads are modeled as constant powers, then the system model would be represented by a set of differential-algebraic equations. In this case, the proposed decomposition and aggregation strategy still applies. In particular, the properties of the voltage subsystem remain largely unchanged (the results are omitted here because of the space limitation), while the properties of the rotor angle subsystem, an implicitly coupled Kuramoto model [42], become more complex and require further study.

**Minquan Chen** received a Ph.D. degree in electrical engineering from the School of Electrical Engineering, Zhejiang University, Hangzhou, China, in 2021. His research interests include power system stability analysis and control.

**Deqiang Gan** (SM'98) has been with the faculty of Zhejiang University since 2002. He visited the University of Hong Kong in 2004, 2005 and 2006. Deqiang worked for ISO New England, Inc. from 1998 to 2002. He held research positions in Ibaraki University, University of Central Florida, and Cornell University from 1994 to 1998. Deqiang received a Ph.D. in Electrical Engineering from Xian Jiaotong University, China, in 1994. He served as an editor for European Transactions on Electric Power (2007-2014). His research interests are power system stability and control.